\documentstyle[eqsecnum,prd,aps,floats,epsfig]{revtex}

\newcommand{\beq}{\begin{equation}}
\newcommand{\eeq}{\end{equation}}
\newcommand{\bea}{\begin{eqnarray}}
\newcommand{\eea}{\end{eqnarray}}

\def\laq{\raise 0.4ex\hbox{$<$}\kern -0.8em\lower 0.62
ex\hbox{$\sim$}}
\def\gaq{\raise 0.4ex\hbox{$>$}\kern -0.7em\lower 0.62
ex\hbox{$\sim$}}

\def\bean{\begin{eqnarray*}}
\def\eean{\end{eqnarray*}}

\def \bk {{\bf k}}
\def \pa {\partial}
\def \ra {\rightarrow}

\def \la {\lambda}

\def \b {\beta}
\def \a {\alpha}

\def \Ga {\Gamma}
\def \ga {\gamma}
\def \sg {\sigma}
\def \da {\delta}
\def \ep {\epsilon}
\def \r {\rho}
\def \om {\omega}
\def \Om {\Omega}

\def \pa {\partial}
\def \dd {\partial}
\def \ra {\rightarrow}

\def \al {\alpha}
\def \la {\lambda}

\def \De {\Delta}
\def \de {\delta}
\def \b {\beta}
\def \a {\alpha}

\def \Ga {\Gamma}
\def \ga {\gamma}
\def \sg {\sigma}
\def \si {\sigma}
\def \Sg {\Sigma}
\def \da {\delta}
\def \ep {\epsilon}
\def \r {\rho}
\def \om {\omega}
\def \Om {\Omega}

\begin{document}
\par
\begingroup
%\twocolumn[%

\begin{flushright}
CERN-TH/98-69\\
gr-qc/9804076\\
\end{flushright}

\vskip 1true cm
{\large\bf\centering\ignorespaces
Seeds of large-scale anisotropy in string cosmology
\vskip2.5pt}
{\dimen0=-\prevdepth \advance\dimen0 by23pt
\nointerlineskip \rm\centering
\vrule height\dimen0 width0pt\relax\ignorespaces
R. Durrer${}^{(1)}$, M. Gasperini${}^{(2)}$,
M. Sakellariadou${}^{(1)}$ and G. Veneziano${}^{(3)}$
\par}
{\small\it\centering\ignorespaces
${}^{(1)}$
D\'epartement de Physique Th\'eorique, Universit\'e de
Gen\`eve, \\
24 quai E. Ansermet,  CH-1211 Geneva, Switzerland \\
${}^{(2)}$
Dipartimento di Fisica Teorica, Universit\`a di Torino, \\
Via P. Giuria 1, 10125 Turin, Italy \\
${}^{(3)}$
Theory Division, CERN, CH-1211 Geneva 23, Switzerland \\
\par}
{\small\rm\centering(\ignorespaces April 1998\unskip)\par}

\par
\bgroup
\leftskip=0.10753\textwidth \rightskip\leftskip
\dimen0=-\prevdepth \advance\dimen0 by17.5pt \nointerlineskip
\small\vrule width 0pt height\dimen0 \relax

%\begin{abstract}
Pre-big bang cosmology predicts tiny
first-order dilaton and metric perturbations at very large scales.
Here we discuss  the possibility that other -- more copiously
generated -- perturbations may act, at second order, as scalar
seeds of large-scale  structure and CMB anisotropies. We study, in
particular, the cases of electromagnetic and axionic seeds. We
compute the stochastic fluctuations of their energy-momentum
tensor and determine the resulting  contributions
to the multipole expansion of the temperature anisotropy.  In the
axion case it is possible to obtain 
a flat or slightly tilted blue
spectrum that fits present  data consistently, both
for massless and for massive (but very light) axions.
%\end{abstract}

\par\egroup
%\vskip2pc]
\thispagestyle{plain}
\endgroup

%\pacs{}

\section {Introduction}
\label{I}

String theory has recently motivated the study of a cosmological
scenario in which the universe, starting from the string
perturbative vacuum, evolves through an early inflationary
``pre-big bang"  phase  \cite{1}, until a transition to the
radiation-dominated, decelerated evolution occurs.

In spite of some attractive aspects of the pre-big bang picture, such
as the underlying duality symmetry \cite{2}, which naturally selects
perturbative initial conditions and automatically leads to inflation
\cite{1,G}, it is
fair to say that such a cosmological scenario is far from being
understood in all of its aspects. In particular, on the more theoretical
side, one is lacking a complete and consistent
description of the high-curvature, strong coupling regime, where
the transition from the pre- to the post-big bang era is expected to  
take place \cite{3}. Furthermore, opinions vary \cite {G,3a}
as to whether or not the pre-big bang
scenario needs a large amount of fine-tuning.
On a more phenomenological side, the main outstanding problem is
to reproduce the observed amplitude and slope
of the large-scale temperature anisotropy \cite{4} and of 
large-scale density perturbations. The difficulty  is that,
unlike in the more conventional (de-Sitter-like) inflationary picture,
the amplification of scalar and tensor metric perturbations
here leads   to  primordial spectra that grow with frequency
\cite{5b}, and whose energy density is normalized to
an almost critical value at some short scale \cite{4a} (typically the
GHz); in this way, too little power is left at scales that
are relevant for anisotropies in the Cosmic Microwave Background
(CMB)~\cite{4} or to the problem of
large-scale structure (unless the high-curvature phase is long
enough and characterized by an almost constant dilaton field
\cite{5a}).

In this paper we  address this problem and we discuss a
possible solution, based on the contribution of ``seeds"  \cite{d90} to
density fluctuations and to the anisotropy of the CMB radiation. The
seeds are produced, in our context, by the amplification of  quantum
fluctuations of some other fields, which are present in string theory,
but are not part of the homogeneous background whose
perturbations we wish to study. 

We shall consider two examples, in which the seed inhomogeneity
spectrum is due, respectively, to  vacuum fluctuations of the
electromagnetic (EM) \cite{7} and of the (Kalb-Ramond) axion
(AX)~\cite{8} field.
Both cases are typical of string cosmology, since no inhomogeneity is
produced, in either case, in a conventional scenario based on
Einstein's equations, without axion and dilaton. The spectra
of EM and AX perturbations can be much flatter than those of scalar
and tensor perturbations of the metric and of the dilaton field.

The idea of using the EM fluctuations as seeds was already 
discussed in a previous paper \cite{mgiovan}, using however the 
perfect fluid approximation for the EM stress tensor. Here we 
will compute the scalar components of the
energy-momentum-tensor fluctuations
due to the EM and AX seeds
including an important anisotropic stress term, and will 
relate them to the primordial spectral energy distributions.
When these seed inhomogeneities are inserted
in the perturbed Einstein equations they generate scalar-metric
fluctuations which are largely controlled, for seeds with small
enough  anisotropic stresses, on super-horizon scales,
 by the so-called  compensation mechanism \cite{mairiandruth}.
Finally,  scalar-metric perturbations can be converted in a standard
manner into temperature fluctuations $\Delta T/T$ via the
Sachs-Wolfe effect \cite{SW}. We will discuss whether the metric
perturbation spectrum induced by seeds can be flat enough to match
present observations, consistently with the COBE normalization of
the amplitude on large scales, and with the high-frequency
normalization of the primordial seed spectrum.

The paper aims at being rather self-contained and readable by
non-specialists in string and/or cosmological perturbation theory,
and is organized as follows. In Section \ref{II} we set up the relevant
equations needed to study super-horizon perturbations in the
presence of seeds, and give their generic solution
for seeds  with ``small" or ``large" anisotropic stresses.
We also discuss the way the perturbations enter the
multipole expansion of $\Delta T/T$ via the Sachs-Wolfe effect.
In Section \ref{III}, after recalling known results about scalar, tensor,
electromagnetic and axion perturbations in the pre-big bang
scenario, we  estimate the contribution of the two latter sources to
the fluctuations of the
energy-momentum tensor, including the case of massive axions. In
Section \ref{IV} we combine the results of the previous two sections and
compute the contribution of EM and AX seeds to $\Delta T/T$. Using
COBE data, we finally  discuss, in the various cases, whether the
seed mechanism alone is able to give a satisfactory explanation of
large-scale temperature anisotropies.  Section \ref{V} contains our
conclusions. Some technical details  are relegated
to the three appendices.
\vspace{0.2cm}\\
{\bf Notation:} The Friedmann metric is given by
$a^2(-d\eta^2+\ga_{ij}dx^idx^j)$, where $a$  denotes the scale
factor and $\eta$ is conformal time. Spatial indices, $1,2,3$ are
denoted by latin letters while spacetime indices, $0,1,2,3$ are
denoted by greek letters. A dot denotes derivative with respect to
$\eta$.

\section {Large-scale perturbations in the presence of seeds}
\label{II}

Before calculating CMB anisotropies for specific examples in the
context of string cosmology, we derive a general formula for
large-scale  CMB anisotropies in models with seed perturbations.

\subsection{Cosmological Perturbation Theory with Seeds}
\label{II1}

In this subsection we give a brief reminder of gauge-invariant
perturbation theory with seeds. More details can be found in
Refs. \cite{d90,review}. By seeds we mean an inhomogeneously
distributed form of energy, which contributes only a small fraction to
the total energy density of the universe and can thus be considered
as a perturbation. Furthermore, we consider seeds that interact
only  gravitationally with the cosmic fluid.

We restrict our discussion to scalar perturbations, which are of
primary interest here. The corresponding equations for
vector and tensor perturbations can be found in \cite{review}. The
metric of a perturbed Friedmann universe is
\beq
g_{\mu\nu} = g^{(0)}_{\mu\nu} + a^2h_{\mu\nu} \; ,
\eeq
where $g^{(0)}$ denotes the unperturbed metric:
\beq
g^{(0)}_{\mu\nu}dx^\mu
dx^\nu=a^2(\eta)(-d\eta^2+\ga_{ij}dx^idx^j)~.
\eeq
Here $a$ is the scale factor, $\eta$ denotes conformal time and $\ga$
represents a metric of constant curvature $K=\pm1,0$. Since we will
be interested in a Friedmann universe that has undergone
substantial inflation, we neglect $K$ in the sequel, setting
$\ga_{ij}=\de_{ij}$.

For scalar perturbations, a Fourier component of $h_{\mu\nu}$ with
wave vector $\bf k$ can by parametrized by 4 scalar
functions $A, B,
H_L$ and $H_T$, defined by
\bea
&&
h(k) = h_{\mu\nu}(k) dx^\mu dx^\nu =
-2A(k)(d\eta)^2 -
 2i{k_j\over k}B(k)d\eta dx^j \nonumber \\
&&
+	2\left[H_L(k) + {1\over 3}H_T(k)\right]\de_{lj}dx^ldx^j -
2{k_lk_j\over
k^2}H_T(k)dx^ldx^j \;.
 \label{2h}
\eea
These four functions  are gauge-dependent, {\em i.e.}
they depend on the choice of coordinates. In order to define
gauge-independent metric variables, we first make use of two
 geometric quantities: the spatial part of the scalar
curvature of the perturbed metric, $\de R$, and the shear (traceless)
part of the extrinsic curvature, $K^{(aniso)}$. An elementary
calculation gives \cite{review}:
 \beq
\de R = 4k^2a^{-2}{\cal R} \mbox{ ,
\hspace{1.2cm} ~~~ }
	{\cal R} = H_L + {1\over 3}H_T  \label{2R} \; ,  \eeq
\beq  K^{(aniso)}_{ij} = ak\left({k_ik_j\over k^2}-{1\over
3}\de_{ij}\right)\si
	\mbox{ , \hspace{1.2cm}~~~  }
    \si = \dot{H_T}/k - B \; .\label{2sigma}
\eeq
Studying the gauge transformation properties of
$A,\; {\cal R},$ and $\si$, one easily finds that the following
variables, called the (Fourier components of the) Bardeen potentials,
are gauge-invariant (see \cite{Bardeen,KS}):
\beq
\Phi = {\cal R} - (\dot{a}/a)k^{-1}\si ~,   \label{2Phi} \eeq
\beq \Psi =  A  - (\dot{a}/a)k^{-1}\si -k^{-1}\dot{\si} \; .
\label{2Psi}
\eeq
(Note that, throughout this paper, we shall always express the
Bardeen potentials in momentum space, even without indicating
their $k$ dependence explicitly.)

Next, we discuss the
perturbations of the energy-momentum tensor.
Let us define the perturbed energy density $\rho^{(pert)}$ and
4-velocity field $u$ as the time-like eigenvalue and eigenvector of
the energy-momentum tensor:
\beq
 T_{\mu}^{\;\;\nu}u^{\mu} =
-\rho^{(pert)} u^{\nu} \;\;,\;\;~~~~~ u^2 = -1 \; .
\eeq
The Fourier components
of the perturbations in the density and velocity field are determined
by
\beq
\rho^{(pert)} = \rho(1+\de)  \;\;, \label{2de} \eeq
\beq  u^0 = (1-A) \;,\; ~~~~~~{u^j\over u^0} = -i{k^j\over k}v \; ,
    \label{2v}
\eeq
where $\rho$ denotes the unperturbed background density. The
 temporal component $u^0$ is   fixed by the normalization
condition. We project  the stress tensor onto the 3-space
orthogonal to $u$:
\beq
\tau_{\mu\nu} =  P_{\mu}^{\rho}P_{\nu}^{\;\;\sigma}
T_{\rho\sigma}, \;\;~~~~~~
P_{\mu\nu} \equiv g_{\mu\nu} + u_{\mu}u_{\nu},
\eeq
and define the scalar perturbations of $\tau$  by:
\beq
\tau_i^{\;j} = p\left[\left(1+\pi_L +{1\over 3}\pi_T\right)\de_i^{\;j}
                      -{k_ik^j\over k^2}\pi_T \right ] \;. \label{2pi}\eeq
The variable $\pi_L$ describes the pressure perturbation, $\pi_T$
is the potential of the anisotropic stresses and $p$ is the
unperturbed background pressure.
Studying the behaviour of the quantities $\de$, $v$, $\pi_L $
and $\pi_T$ under gauge transformations \cite{DS}, one finds the
gauge-invariant variables:
\bea
&&
  \Pi =\pi_T ,  ~~~
  \Ga = \pi_L - (c_s^2/w)\de ,  ~~~
   V = v - k^{-1}\dot{H_T}, \nonumber     \\
 && D = \de +3(1+w)(\dot{a}/a)k^{-1}(V + \si), ~~~
  D_g =\de +3(1+w){\cal R}~.
      \eea
Here $\Pi$ is the anisotropic stress potential, $\Gamma$ is the
entropy perturbation, $V$ is the peculiar velocity potential, $D$ and
$D_g$ are different choices for a gauge-invariant density
perturbation variable (for a physical interpretation of these
variables, see \cite{KS,DS}).  Finally, $w=p/\rho$ denotes the enthalpy
and $c_s^2=\dot{p}/\dot{\rho}$ stands for the adiabatic  speed of
sound. In this paper we shall limit ourselves to adiabatic
perturbations ($\Gamma=0$).

The perturbation of Einstein's equations and of
energy-momentum conservation can be expressed in
terms of these gauge-invariant variables (a
derivation  can be found in \cite{KS,DS}). We obtain two
 constraint equations:
\bea
  4\pi Ga^2\rho D &=& k^2\Phi , \label{2C1}  \\
  4\pi Ga^2(\rho +p)V &=& k\left[(\dot{a}/ a)\Psi -\dot{\Phi}\right];
\label{2C2}
 \eea
two  dynamical equations:
\bea
&& 	
-8\pi Ga^2p\Pi = k^2(\Phi+\Psi)  \label{2D1} , \\
&&
8\pi Ga^2p\left[\Ga + (c_s^2/w)D_g +(2/3)k^2\Pi\right]=
\nonumber\\
 && =	{\dot{a}\over a}\left\{\dot{\Psi}-
\left[a^{-1}\left(\frac{a^2\Phi}{\dot{a}}\right)^{\bullet}
\right]^{\bullet}\right\}
+ \left[{2\over a}\left({\dot{a}\over a^2}\right)^{\bullet}+
3\left({\dot{a}\over a^2}\right)^2\right]
\left	[\Psi-a^{-1}\left(\frac{a^2\Phi}{\dot{a}}\right)^{\bullet}\right];
\label{2D2}
\eea
and two conservation equations:
\bea
  \dot{D_{\al}} -3w_{\al}(\dot{a}/a)D_{\al}  &=&
 -k\left[(1+w_{\al})V_{\al} +2(\dot{a}/a)w_{\al}k^{-1}\Pi_{\al}\right]
  \nonumber   \\ &&+
  3(1+w_{\al})4\pi Ga^2(\rho +p)(V-V_{\al})
  \; , \label{2con1}\\
\dot{V}_{\al} +(\dot{a}/a)V_{\al} &=& \frac{c^2_{\al}}{1+w_{\al}}
kD_{\al} +
\frac{w_{\al}}{1+w_{\al}}k\Ga_{\al} + k\Psi -
\frac{2w_{\al}}{3(1+w_{\al})}k\Pi_{\al}  \; . \label{2con2}
\eea
The above conservation equations  hold for any component $\a$ of the
fluid stress-energy tensor which interacts with the other components
of the cosmic fluid only  gravitationally.  The variables $c_{\al}$ and
$w_{\al}$ denote the adiabatic speed of sound and the enthalpy  of
the fluid component,  respectively. The total perturbations are
defined as the sums:
\beq
\rho D = \sum_{\al}\rho_{\al}D_{\al}  \;\;,
\;\;\;
	(\rho +p)V = \sum_{\al}
  	(\rho_{\al} + p_{\al})V_{\al} \;\;, \; \mbox{ etc.}
\eeq
For interacting matter, the corresponding equations  can be found
  in \cite{KS}.

In order to complete the above analysis
we also need equations of state for the matter sources,
which  relate for instance
$\Ga$ and $\Pi$ to $D$ and $V$. Due to the Bianchi identities,
the conservation equations for the total cosmic fluid follow from the
field equations (\ref{2C1})--(\ref{2D2}). Thus, we  need not make
explicit use of both dynamical equations,  but we can
use,  say, (\ref{2D1}) and one of
the conservation equations (\ref{2con1}), (\ref{2con2})
for the total fluid.

We now add to the perturbation equations an inhomogeneous
energy-momentum distribution, $T_{\mu\nu}^{(s)}$,
generated by seed fields that
do not interact with the cosmic fluid other than gravitationally.

Since, by definition,  seeds do not  contribute as
sources of the homogeneous background,
 the energy-momentum tensor  $T_{\mu\nu}^{(s)}$
is gauge-invariant by
itself \cite{StWa}, and
 can be calculated by solving the field
equations for the seeds in the unperturbed  background
geometry.  Let us assume that we can express the Fourier
components of $T_{\mu\nu}^{(s)}$ in terms of four scalar
``seed-functions" $f_\rho$, $f_p$, $f_v$ and $f_\pi$
(we just neglect vector and tensor contributions; since they are
decoupled from  density
perturbations, in the linear approximation, this
will not affect our results for scalar perturbations):
\bea
T_{00}^{(s)}({\bf k},\eta) &=& a^2\rho^{(s)} =
	M^2f_{\rho}({\bf k},\eta) \; ,
     \label{3seed00} \\
     T_{j0}^{(s)}({\bf k},\eta)  &=& -i{k_j\over k}a^2v^{(s)}
	= -iM^2k_jf_{v}({\bf k},\eta) \; ,
     \label{3seed0i} \\
   T_{ij}^{(s)}({\bf k},\eta)    &=& a^2\left[\left(p^{(s)} +{1\over  
3}\Pi^{(s)}\right)\ga_{ij}
                       -{k_ik_j\over k^2}\Pi^{(s)}\right]   \nonumber \\
    &=& M^2\left
[\left(f_p ({\bf k},\eta)+{k^2\over 3}f_{\pi}({\bf k},\eta)
\right)\ga_{ij} -k_ik_jf_{\pi}({\bf k},\eta)\right] \;.
 \label{3seedij}
\eea
Note that  $f_\rho$ and $f_p$ have dimension $\ell^{-2}$, while
$f_v$ has dimension $\ell^{-1}$ and $f_\pi$ is dimensionless.
Here $M$ denotes an arbitrary mass scale, introduced for dimensional
reasons, which will eventually drop out in physical predictions.

Given an energy-momentum tensor $T_{\mu\nu}$, which
in general contains vector and tensor contributions, the scalar parts
$f_v$ and $f_{\pi}$ are  determined by the identities:
\bea
ik^jT^{(s)}_{0j}& =& M^2k^2 f_v  , \nonumber \\
 -T^{(s)}_{ij}k^ik^j +{1\over 3}k^2\ga^{kl}T^{(s)}_{kl}&=&
   \frac{2}{3}M^2k^4f_{\pi}   \; .
\eea
On the other hand, $f_v$ and $f_{\pi}$ are related to  $f_{\rho}$
and $f_p$, by the conservation equations
$\nabla^\nu T^{(s)}_{\mu\nu}=0$:
\beq
\dot{f}_{\rho} + k^2f_v + (\dot{a}/a)(f_{\rho} +3f_p) = 0,
\label{0f}\eeq
\beq \dot{f}_v +2 (\dot{a}/a)f_v -f_p + (2/3)k^2f_{\pi} = 0~.
 \label{3f}  \eeq

In the presence of seeds, and in the approximation in which
perturbations are treated linearly,
the total geometric perturbations can be separated into a part
induced by the seeds, $\Psi_s ,\Phi_s$,
 and a part induced by the perturbations of the
cosmic fluid, $\Psi_m ,\Phi_m$. The perturbed Einstein's equations
 (\ref{2C1}) and (\ref{2D1}) become
\bea
 k^2\Phi &=& 4\pi G\rho a^2 D+ \ep\left[f_{\rho} +
3(\dot{a}/a)f_v\right]
       \; , \label{2G1} \\
 \Phi +\Psi &=&-8\pi G a^2k^{-2} p\Pi -2\ep f_{\pi} \; , \label{2G3}
\eea
where $\ep \equiv 4\pi GM^2$. If we define
\beq
\Psi = \Psi_s + \Psi_m, \,\,\,\,\,\,\,\,  \Phi = \Phi_s + \Phi_m
\label{defsm}
\eeq
with:
\beq
k^2 \Phi_s = \ep\left[f_{\rho} + 3 (\dot{a}/a)f_v)\right],
\,\,\,\,\,\,\,\,
\Phi_s + \Psi_s = - 2 \ep f_{\pi},
\eeq
we easily find
\bea
\Phi_m
    &=&4\pi G\r a^2 k^{-2}\left[D_g+3(1+w)\left({\dot a\over a}\right)
{V\over k}-3(1+w)\Phi \right] ,
	\label{phim}\\
\Psi_m&=&-\Phi_m-8\pi G a^2 p \Pi k^{-2}~.	\label{psim}
\eea
 Equation~(\ref{phim}) has been written in terms of the
gauge-invariant density perturbation  $D_g$, because this choice will
simplify our final equations. Physically, $D_g$ corresponds to the
density perturbation in the flat slicing. The evolution of $D_g$ and
$V$ is described by the conservation equations (\ref{2con1}) and
(\ref{2con2}), which read explicitly:
\bea
\dot {D_g} +3(c_s^2-w){\dot a\over a}D_g&=&-(1+w)kV
, \label{Dgdot}\\ \dot V+{\dot a\over
a}(1-3c_s^2)V&=&k(\Psi-3c_s^2\Phi)+ k{c_s^2\over 1+w}D_g
-{2w\over 3(1+w)}k\Pi~. \label{d01}
\eea

To simplify the analysis, we will assume $w=c_s^2=$ constant. The
unperturbed background equations are then solved by $a\propto
\eta^r$, with $r=2/(3w+1)$. Since we are interested in very large
scale perturbations in the cosmic microwave background,  we
concentrate our discussion on super-horizon scales, such that $
k\eta\ll 1$. Eqs.~(\ref{phim}) and (\ref{defsm}) then
lead  to
\beq
\Phi = {1\over 3(1+w)}D_g +{r\over k\eta}V+{2\over
9r^2(1+w)}(k\eta)^2\Phi_s ~, \label{f1}
\eeq
where $r=1$ for the radiation-dominated era, and $r=2$ for the
matter-dominated era.

The evolution equation for $D_g$, Eq.~(\ref{Dgdot}), implies
$dD_g/d (k\eta) = -(1+w)V$. In the
 physical picture we have in mind, metric perturbations are
triggered by the presence of the seeds alone, and we do not want to
include an arbitrary contribution from the perturbations of the
homogeneous sources. We thus
 require $D_g(0)=0$, which implies
$D_g\sim k\eta V$. Hence, we may  neglect the $D_g$-term in
Eq.~(\ref{f1}) for $k\eta\ll 1$.

 Combining Eqs.~(\ref{f1}), (\ref{d01}),
(\ref{2G3}) we  find, on super-horizon scales,
\beq
\Psi = {dV\over d (k\eta)} + {r\over k\eta}V + {2 w  \over 3
r^2(1+w)}(k\eta)^2\Phi_s
	+{2 (k\eta)^2 \over 9r^2(1+w)} (2 \ep f_{\pi}+\Psi+\Phi) \;.
\label{psi}
\eeq

The two equations (\ref{f1}) and (\ref{psi})
relate the three variables $\Psi, \Phi$ and
$V$ once the seeds are given. To proceed, we need an equation
of state to close the system. For single  component fluids
 this equation usually takes the form
$\Pi=\Pi(D_g,V)$. We are interested in large-scale CMB anisotropies,
which are induced at
 recombination and later, when  the universe is already
matter-dominated, with $p\ll\rho$. Thus, in what follows, we will
consider the case  $\Pi=0$, which implies
\beq
 \Phi+\Psi=-2\ep f_\pi.
\label{PhiPsi}
\eeq
Furthermore, in a  matter-dominated  Friedmann universe, $r=2$
and $w=0$. The equation of motion for $V$, obtained by combining
Eqs.~(\ref{f1}), (\ref{psi}), (\ref{PhiPsi}),  then reads
\beq
{dV\over d (k\eta)} +{4\over k\eta}V=-{1\over
18}(k\eta)^2\Phi_s-2\ep f_\pi ~ = -{1\over 18} \eta^2
\ep\left[f_{\rho} + 3 (\dot{a}/a)f_v\right]-2\ep f_\pi ~.
\label{Vprime}
\eeq
In the next subsection, we shall see that the large-scale anisotropies
of the CMB are determined by the combination $\Psi$--$\Phi$. Using
Eqs.~(\ref{f1}), (\ref{PhiPsi}) and (\ref{Vprime}), we find
immediately:
\beq
\Psi{\rm -}\Phi ={dV\over d (k\eta)} -{1\over 18}
\eta^2 \ep\left[f_{\rho} +
3 (\dot{a}/a)f_v\right]~. \label{Psi-Phi}
\eeq
Modulo numbers of order unity, which can be computed case by case,
 we finally arrive at the estimates:
\beq
\Psi - \Phi  \sim {dV\over d (k\eta)} \sim {V\over k\eta} \sim \max
\left\{\ep f_\pi, \ep \eta^2\left(f_\r+3{\dot a \over a}f_v
\right)\right\}~. \label{comp}
\eeq
Depending on whether $\eta^2 \left(f_\r+3(\dot a /a)f_v\right)$
or $f_\pi$  dominates in Eq.~(\ref{comp}),  we can
distinguish between seeds with small and large anisotropic
stresses. We will discuss in Section \ref{III} to which case
 our string-cosmology seeds belong.

If the term $\ep\eta^2 \left(f_\r+3(\dot a / a)f_v\right)=
x^2\Phi_s$ dominates, we conclude from
Eqs.~(\ref{Vprime}),(\ref{Psi-Phi}) that \beq
\Phi \sim \Psi \sim (k\eta)^2\Phi_s \sim (k\eta)^2\Psi_s \ll
\Phi_s\sim\Psi_s ~, \eeq
on super-horizon scales.
This suppression of the total geometric perturbations, if compared with
the source perturbations alone, is known under the name of 
``compensation''\cite{mairiandruth}. The conservation equations
(\ref{2con1}), (\ref{2con2}) show that the presence of seeds induces
matter perturbations that try to compensate the gravitational
potential of the seeds. Since anisotropic stresses in the seeds
cannot be compensated by a perfect fluid, compensation is not
effective, if anisotropic stresses
 dominate.  But, as shown here (see also
\cite{mairiandruth}), the phenomenon of compensation is quite
generic and, to a large extent,  independent of the spectrum of seed
fluctuations.

\subsection{ The Seed Contribution to CMB anisotropies}
\label{II2}

In this subsection we calculate the CMB anisotropies for models
where perturbations are induced by seeds, and their contribution to
$\Delta T/T$ via the Sachs-Wolfe effect \cite{SW}. We first discuss in
general the motion of photons in a perturbed Friedmann universe.

We make use of the fact that the equations of motion of
photons are conformally invariant. More precisely, two metrics that
are conformally equivalent,
\beq d\bar{s}^2 = a^2ds^2 \; ,
\eeq
have
the same light-like geodesics, only the corresponding affine
parameters  are different. Let us denote the two affine parameters
by $\bar{\la}$ and $\la$ respectively, and the tangent vectors to the
geodesic by
\beq n = \frac{dx}{d\la}, \,\,\,\, \;  \bar{n} =
\frac{dx}{d\bar{\la}} \;\;, \;\;\; n^2 = \bar{n}^2 = 0 \;, \;\;  n^0 =1
\;,\;\; {\bf n}^2 =1.
\eeq
Setting $n^0 = 1 +\de n^0$, the geodesic
equation for the perturbed metric
\beq ds^2 =
(\eta_{\mu\nu}+h_{\mu\nu})dx^{\mu}dx^{\nu}  \eeq
 yields, to first
order,
\beq \de n^0 |_i^f = \left[h_{00} + h_{0j}n^j\right]_i^f -
   {1\over 2}\int_i^f\dot{h}_{\mu\nu}n^{\mu}n^{\nu}d\la  \; .
\label{2deltan}
\eeq
On the other hand, the ratio of the energy of a photon measured by
some observer at $t_f$ to the energy emitted at $t_i$ is
\beq
{E_f\over E_i} = \frac{(\bar{n}\cdot u)_f}{(\bar{n}\cdot u)_i}
	= {T_f\over T_i}
     \frac{(n\cdot u)_f}{(n\cdot u)_i}  \; , \label{Ef/Ei}
\eeq
where $u_f$ and $u_i$ are the four-velocities of the observer and
emitter
respectively, and the factor $T_f/T_i$ is the usual (unperturbed)
redshift, which relates
 $n$ and $\bar{n}$.
The velocity field of observer and emitter is given by
 \beq u = (1-A)\dd_\eta +v^i\dd_i \; . \eeq

An observer measuring
a temperature  $T_0$ receives photons that were emitted at the
time $\eta_{dec}$ of decoupling of matter and radiation, at the fixed
temperature $T_{dec}$. In first-order perturbation theory, we find the
following relation between the unperturbed temperatures $T_f$,
$T_i$,  the measurable temperatures $T_0$, $T_{dec}$, and the photon
density perturbation:
\beq {T_f \over T_i} =
	{T_0\over T_{dec}}\left(1 - {\de T_f\over T_f} + {\de T_i
\over T_i}\right) =
    {T_0\over T_{dec}}\left(1 - {1\over 4}\de^{(\ga)}|_i^f\right) \; ,
\eeq
where $\de^{(\ga)}$ is the intrinsic density perturbation in the
radiation and we used $\rho^{(\ga)}\propto T^4$ in the last
equality. Inserting the above equation 
and Eq.~(\ref{2deltan}) into
Eq.~(\ref{Ef/Ei}),
and  using Eq.~(\ref{2h}) for the definition of $h_{\mu\nu}$,
one finds, after  integration by parts \cite{review}:
\beq
 {E_f\over E_i} = {T_0\over T_{dec}}\left\{1-\left[ {1\over
4}D^{(\ga)}_g +
	  V_j^{(m)}n^j  +\Psi-\Phi\right]_i^f +
   	\int_i^f(\dot{\Psi}-\dot{\Phi})d\la\right\}  \; .
\label{2deltaE}  \eeq
Here $D_g^{(\ga)}$ denotes the density perturbation in the radiation
fluid, and  $V^{(m)}$ is the peculiar velocity of the baryonic matter
component (the emitter and observer of radiation).
The final time values in the square bracket of Eq. (\ref{2deltaE}) give
rise only to monopole contributions and to the dipole due to our
motion with respect to the CMB, and will be neglected in what
follows.

Evaluating Eq.~(\ref{2deltaE}) at final time $\eta_0$ (today) and
initial time $\eta_{dec}$, we obtain the temperature difference
of photons coming from different directions $\bf n$ and ${\bf n}'$
 \beq {\De T\over T} \equiv
{\de T({\bf n})\over T}- {\de T({\bf n}')\over T},
\eeq
with  temperature perturbation
\beq
{\delta T({\bf n})\over T} =\left[ {1\over 4}D^{(\ga)}_g +
V_{j}^{(m)}n^j
+\Psi -\Phi\right](\eta_{dec},{\bf x}_{dec})
   + \int_{\eta_{dec}}^{\eta_0}(\dot{\Psi}-\dot{\Phi})(\eta,{\bf
	x}(\eta))d\eta~, \label{dT0}
\eeq
where ${\bf x}(\eta)={\bf x}_0-(\eta_0-\eta){\bf n}$ is
the unperturbed photon position at time $\eta$ for an observer at
${\bf x}_0$, and ${\bf x}_{dec}={\bf x}(\eta_{dec})$. The first
term in Eq.~(\ref{dT0}) describes the intrinsic
inhomogeneities on the surface of the last scattering, due to acoustic
oscillations prior to decoupling. In general, it also contains
contributions  to the geometrical perturbations.  This is especially
important in the case of adiabatic inflationary models \cite{ram}.
However, for perturbations induced by seeds, which satisfy the
initial condition $D_g(k,\eta)\ra 0$ for ${\eta\ra 0}$, the geometrical
contributions to $D_g$ can be neglected. The second term describes
the relative motions of emitter and  observer. This is the Doppler
contribution to the  CMB anisotropies. It appears on the same
angular scales as the acoustic term, and we thus call the sum of
the acoustic and Doppler contributions ``acoustic peaks''.

The last two terms are due to the inhomogeneities in the spacetime
geometry; the first contribution determines the change in the photon
energy due to the difference of the gravitational potential at the
position of emitter and observer. Together with the part contained in
$D_g^{(r)}$ they represent the ``ordinary'' Sachs-Wolfe  effect. The
second term accounts for red-shift or blue-shift caused by the
time dependence of the gravitational field along the  path of the
photon, and represents the so-called Integrated Sachs-Wolfe (ISW)
effect. The sum of the two terms is the full  Sachs-Wolfe contribution
(SW).

On angular scales
$0.1^\circ\stackrel{<}{\sim}  \theta\stackrel{<}{\sim}  2^\circ$, the
main contribution to the CMB anisotropies comes from the acoustic
peaks, while the SW effect is  dominant  on large angular scales. On
scales smaller than about $0.1^\circ$, the anisotropies are damped
by the finite thickness of the recombination shell, as well as by
photon diffusion during recombination (Silk damping). Baryons and
photons are very tightly coupled before recombination, and oscillate
as a one-component fluid. During the process of decoupling, photons
slowly diffuse out of over-dense regions into under-dense ones. To fully
account for this process, one has to solve the Boltzmann equation
for the photons (see, e.g. \cite{review}).

The angular power spectrum of
CMB anisotropies is expressed in terms
of the dimensionless coefficients $C_\ell$, which appear in the
expansion of the angular  correlation function in terms of the
Legendre polynomials $P_\ell$:
\beq
\left\langle{\delta T\over T}({\bf
n}){\delta T\over T}({\bf n}') \right\rangle_{{~}_{\!\!({\bf n\cdot
n}'=\cos\vartheta)}}=
  {1\over 4\pi}\sum_\ell(2\ell+1)C_\ell P_\ell(\cos\vartheta)~.
\label{cor}
\eeq
Here the brackets denote spatial average, or expectation values if
perturbations are quantized.

To determine  the $C_{\ell}$ we Fourier-transform Eq.~(\ref{dT0}),
defining
\beq
\varphi({\bf k}) = {1\over \sqrt{V}}\int_V
	\varphi({\bf x})e^{i{\bf k\cdot x}}d^3x~,
\eeq
 and using the identity
\beq
e^{iz\cos\vartheta}=\sum_\ell (2\ell +1)i^{\ell}j_{\ell}(z)P_{\ell}
(\cos\vartheta)\nonumber
\eeq
(where $j_{\ell}$ is the spherical Bessel function of order $\ell$).
For the
coefficients $C_\ell$ of Eq.~(\ref{cor}) we obtain:
\begin{equation}
C_\ell = {2\over \pi} \int
{\langle|\Delta_\ell ({\bf k})|^2\rangle  \over (2\ell +1)^2} k^2 dk ~,
\end{equation}
where
\begin{eqnarray}
{\Delta_\ell \over 2\ell +1} &=&j_\ell(k\eta_0)\left[
	{1\over 4}D_g^{(r)}({\bf k},\eta_{dec})
	+(\Psi -\Phi)({\bf k}, \eta_{dec})\right]
	- j_\ell '(k\eta_0) {\bf V}_r({\bf k},\eta_{dec})
\nonumber \\
&&+\int_{\eta_{dec}}^{\eta_0}(\dot{\Psi}-\dot{\Phi})({\bf k},\eta')
j_\ell \left(k\eta_0-k\eta'\right) d\eta'
\nonumber \\
&=&{1\over 4}D_g^{(r)}({\bf k},\eta_{dec})j_\ell(k\eta_0)
- j_\ell '(k\eta_0){\bf V}_r({\bf k},\eta_{dec})
\nonumber \\
&& + k\int_{\eta_{dec}}^{\eta_0}(\Psi -\Phi)({\bf k},\eta')
j_\ell '\left(k\eta_0-k\eta'\right) d\eta'~,
\label{Dl}
\end{eqnarray}
and $j'_\ell$ stands for the derivative of $j_{\ell}$ with respect to
its argument. On large angular scales, $k\eta_{dec}\ll 1$ (which
corresponds to $\ell \ll 100$), the SW contribution dominates:
\beq
C_\ell^{SW} = {2\over\pi}\int
k^4dk\left\langle\left[\int_{\eta_{dec}}^{\eta_0}(\Psi -\Phi)({\bf k},
\eta)j_{\ell}'\left(k\eta_0-k\eta\right) d\eta\right]^2\right\rangle  .
\label{Cell}
 \eeq

Let us approximate the Bardeen potentials on super-horizon scales
by a power-law spectrum:
 \beq
 \langle|\Psi-\Phi|^2\rangle= C^2(k)~(k\eta)^{2\ga} ~
\label{power-law}.
\eeq
Furthermore, we consider models where the seed contribution
does not grow in time on sub-horizon scales. In this case the
Bardeen potentials, inside the horizon,
are dominated by the cold dark matter
contribution, which leads to time-independent $\Phi$ and $\Psi$.
 We can thus approximate the Bardeen potentials by
\beq
 \Psi-\Phi \approx \left\{\begin{array}{ll}
	C(k)(k\eta)^\ga & ~,~ k\eta\ll 1\\
	C(k) & ~,~ k\eta\gg 1~.	\end{array}  \right.
\label{252}
\eeq
We further assume that also $C(k)$ is given by a simple power law.
Thus, for dimensional reasons, it  has the form
\beq
C(k)=\left\{\begin{array}{ll}
	{\cal N}k^{-3/2} (k/k_1)^{\al} ~,& k\le k_1\\
	0 ~,& k> k_1 ~,\end{array} \right.
\label{253}
\eeq
where  ${\cal N}$ is a dimensionless constant, and $k_1$ denotes a
comoving  cutoff scale, i.e. the maximal amplified frequency
determined by the explicit mechanism of seed
production (in the case $\alpha=0$ no cutoff is needed).
Inserting this in Eq.~(\ref{Cell}),
\beq
C_\ell^{SW} \approx {\cal N}^2{2\over \pi}\int_0^{k_1} {dk\over
k}
    \left({k\over k_1}\right)^{2\al}|I(k)|^2 	
	~,  \label{2.52}
\eeq
where, setting $x=k\eta, x_0=k\eta_0, x_{dec}=k\eta_{dec}$,
\bea
I(k)   &=&   \int_{x_{dec}}^1dxx^{\ga}j'_\ell(x_0-x)
	+\int_1^{x_0}dxj'_\ell(x_0-x)  \\
  &=&   \int_{x_{dec}}^1dxx^{\ga}j'_\ell(x_0-x) + j_\ell(x_0-1) ~.
\label{Intx}
\eea
We can see explicitely from this equation that the relevant
contribution of each mode to the CMB anisotropy
 comes while the mode is
still outside the horizon ($k \eta <1$). We now distinguish two cases.

If $\ga > -1$ the lower bound  in eq.
(\ref{Intx}) can be safely extended to $0$, and the integral is
dominated by the region $k \eta \sim 1$, so that:
\beq
I(k) \sim j_{\ell}(x_0-x_{dec})\sim j_{\ell}(x_0),~~ ~~
x_{dec}\ll 1 < x_0~.
\eeq
Inserting this in Eq.~(\ref{2.52}), the integral can
be performed exactly (assuming $\eta_0k_1\gg \ell$),  with the
result, for $\a<1$,
\beq
C_{\ell}^{SW} \approx {\cal N}^2(k_1\eta_0)^{-2\al}{\Ga(2-2\al)\over
4^{(1-\al)}\Ga(3/2-\al)} {\Ga(\ell+\al)\over\Ga(\ell + 2-\al)}, ~~~~~~~
\a < 1
\label{simple}
\eeq
(if $\a>1$, the integral grows towards large $k$ and is dominated by  
the contributions at $k\sim k_1$, leading to an $\ell$-independent
result of order  $({\cal N}/k_1\eta_0)^2$).
Comparing the above equation with the standard inflationary result
\cite{JamesBond},
\beq
C_\ell^{SW} \propto {\Ga(\ell -1/2+ n/2)\over\Ga(\ell+5/2-n/2)} ~,
	\label{inflat}
\eeq
where $n$ denotes the usual spectral index,
we find that $\a$ is related to $n$ by $\al=(n-1)/2$. 
The scale-invariant spectrum, as it has been observed by the DMR
experiment aboard the COBE satellite \cite{smootscott}, requires
\beq
 0.8\le n\le 1.4 \label{259a}
\eeq
so that,
allowing for  generous error bars, the COBE observations  imply
 \beq
 -0.1\le \al\le 0.2  ~ ,~~~~~~~~~ \ga > -1~ .  \label{alpourga0}
\eeq

Consider now the second case, $\ga+1\le 0$. The integral (\ref{Intx}) is now
 dominated by its value
at the lower boundary and we get
\beq
|I(k)|^2 \approx {1\over (\ga+1)^2}x_{dec}^{2(\ga+1)}\left[
	{\ell \over 2\ell+1}j_{\ell-1}(x_0) -
	{\ell+1 \over 2\ell+1}j_{\ell+1}(x_0)\right]^2 ~. \label{Ik}\eeq
If also $\al+\ga <0$, the $k$-integral converges and we
obtain (see Appendix~A):
\bea
\lefteqn{C_\ell^{SW} \approx} \nonumber \\
 && {{\cal N}^2\over 2^{2(\al+\ga)}(\ga+1)^2}
	{\Ga(-2(\al+\ga))\over\Ga(1/2-(\al+\ga))^2}
 	\left({\eta_{dec}\over \eta_0}\right)^{2(\ga+1)}
	\left(k_1\eta_0\right)^{-2\al}
        {\Ga(\ell+1+\al+\ga)\over \Ga(\ell+1-\al-\ga)}
	{1\over (2\ell+1)^2} \nonumber \\ \nonumber \\
 &&	\times \left[{\ell^2(\ell-\al-\ga)\over \ell+\al+\ga}
        +{2\ell(\ell+1)(1/2+\al+\ga)\over
	(1/2-\al-\ga)} +{(\ell+1)^2
   (\ell+1+\al+\ga) \over (\ell+1-\al-\ga)}  \right] ~.
\label{C_ell_sw}
\eea
Comparing again this last result with that of 
standard inflation, Eq. (\ref{inflat}), 
and  neglecting the weak $\ell$-dependence of
$(2\ell+1)^{-2}[\cdots ]$ in Eq.~(\ref{C_ell_sw}),  we obtain
\beq
 n\sim 3+2(\al+\ga)~, ~~~~~~~
\alpha + \gamma < 0~.
 \label{indexa}
\eeq
(If, on the contrary, $\al+\ga >0$, the coefficients $C_\ell$ are  
dominated by the large $k$ (i.e. small-scale)
contribution,  even for
the very low values of $\ell$.
In this case the small-scale perturbations become too
large, which is excluded observationally by the
fact that the spectrum, for CMB and matter perturbations,
must be close to the  Harrison-Zel'dovich  spectrum \cite{HZ}).

The observational limits on $n$ thus impose
\beq
 -0.1< \ga+1+\al < 0.2 ,~~~~~ \ga \le -1~, ~~~~~n\simeq 3+2(\a+\ga)
\label{273}
\eeq
and
\beq
 -0.1<\al<0.2 ,~~~~~~~ \ga > -1~, ~~~~~n=1+2\a.
\label{index>-1}
\eeq
In the following sections we will apply these findings to 
electromagnetic  and axionic seeds produced in string cosmology. In the
axions case we will  discuss separately massless and massive 
perturbations.

\section {Seeds from string cosmology}
\label{III}

In this section we compute the seed functions $f_\r, f_v, f_\pi$, and
we estimate the Bardeen potentials for electromagnetic and axion
perturbations, including the case of massive axions.

\subsection{Amplification of quantum fluctuations}
\label{III1}

We start by recalling the form of the (string-frame) low-energy
string effective action \cite{22a}:
\beq
\Gamma^{S}_{eff} = \int d^Dx \sqrt{|g|} e^{-\phi}\left(R +
g^{\mu\nu}
 \partial_{\mu} \phi \partial_{\nu} \phi -{1 \over 12}g^{\mu\rho}
g^{\nu\sigma} g^{\alpha\beta}
 H_{\mu\nu\alpha}H_{\rho\sigma\beta} -
{1 \over 4} g^{\mu\nu}g^{\rho\sigma}F_{\mu\rho}F_{\nu\sigma}
\right) ,
 \label{Saction}
\eeq
where we have included the antisymmetric tensor $H_
{\mu\nu\alpha}=
\pa_{[\mu}B_{\nu\a]}$ and  the $U(1)$ gauge field
$F_{\mu\nu}=\pa_{[\mu}A_{\nu]}$.
Note that this gauge field
is typical of what emerges from heterotic string compactification.
For gauge fields originating {\it \`a la} Kaluza-Klein, the action
and the spectra are somewhat different, as discussed in \cite{B2}.

Upon compactification down to four dimensions, and after
introduction
of the axion field $\sigma$ by the duality transformation:
\beq
H^{\mu\nu\alpha} =
e^{-\phi} \epsilon^{\mu\nu\alpha\beta} \partial_{\beta} \sigma,
\label{Baxduality}
\eeq
one easily arrives at the dimensionally reduced action:
\beq
\Gamma^{S}_{eff} = \int d^4x \sqrt{|g|} e^{-\phi}\left(R +
g^{\mu\nu}
 \partial_{\mu} \phi \partial_{\nu} \phi -{1\over 2} e^{2\phi}
g^{\mu\nu} \partial_{\mu} \sigma \partial_{\nu} \sigma -
{1 \over 4} g^{\mu\nu}g^{\rho\sigma}F_{\mu\rho}F_{\nu\sigma}
\right)~.
 \label{redaction}
\eeq

The study of tensor (T), scalar-dilaton (SD),
electromagnetic (EM)
and axion (AX) perturbations is conveniently performed defining the
external  ``pump  field", responsible for their amplification.
To this aim, we first identify for each
perturbation the canonical
variables $\psi^i$, which diagonalize the perturbed action expanded
up to second order \cite{22b}. In a purely metric-dilaton background,
such variables are easily found from (\ref{redaction}) to be:
\bea
\psi^T = a e^{-\phi/2} h^{TT} \equiv a_{E}  h^{TT}, \;\,\,\,\,\,\,
\psi^{SD} = a e^{-\phi/2} \phi + \dots, \; \nonumber \\
\psi^{EM} =  e^{-\phi/2} A_{\mu}, \;\,\,\,\,\,\,
\psi^{AX} = a e^{\phi/2} \sigma \equiv {a_A} \sigma.
\label{can}
\eea
Here $h^{TT}$ denotes the transverse-traceless part
 of the metric perturbations, the dots in the equation
for $\psi^{SD}$ represent the additional scalar-metric terms needed to
reproduce the gauge-invariant scalar perturbation \cite{22b},
$a_E$ is the scale factor in the Einstein frame, and $a_A$ in the axion
frame \cite{8}. By varying the perturbed action, we find that  the
Fourier modes $\psi_k(\eta)$ of each of these four perturbations
satisfy decoupled, linear equations of the type:
\beq
\ddot\psi_k + \left(k^2- {\ddot{P}\over P}\right)
\psi_k=0~,
\label{evoluzione}
\eeq
where $P(\eta)$ is the pump field, obtained for each case
from eq. (\ref{can}) as:
\beq
P^T=P^{SD}=a_E~; ~~~~~~~P^{EM}=e^{-\phi/2}~;~~~~~~~P^{AX}=a_A~.
\eeq

At the beginning of the inflationary era, characterized by an
accelerated evolution of the pump field, every perturbation is well
inside the  horizon and Eq. (\ref{evoluzione}) has oscillating
solutions, which can be consistently normalized to a vacuum
fluctuation spectrum. During the whole
pre-big bang  phase the general solution can be written in
terms of Hankel functions \cite{11}, with a Bessel index
determined by the power that characterizes the background
evolution (in conformal time) of the pump field. This behaviour has
to be matched with the one after the pre-big bang phase  when,  as
we assume, the universe becomes radiation-dominated and the
dilaton freezes to its present value. In all four cases this implies a
free Klein-Gordon equation for the canonical variable after the period
of accelerated evolution. By matching the pre-big bang and
radiation solutions of the perturbation equations,  we eventually
obtain the final amplified perturbations during the radiation era.

For T and SD perturbations the time evolution of the background
leads to a spectrum that is in general too steep \cite {5b} (see also
\cite{21}) to be expected to give any significant contribution to
very large scale structures, or to temperature anisotropies on the
COBE scale. The only way to achieve a reasonable contribution would
be to have a very long  string phase with an almost constant dilaton
\cite{5a},  which is not excluded,
in principle, either theoretically or
phenomenologically, but which looks somewhat unlikely,
from  both points of view.

For EM perturbations, however, the situation seems to be more
interesting. Consider in fact the transition from a pre-big bang
phase, with growing dilaton ($\phi= -2\b \log |\eta|$), to the
standard radiation-dominated phase with $\phi=$ const, and call
$\eta_1$ the transition time scale. The electromagnetic fluctuations
are directly coupled to the dilaton background, in such a way that
each polarization mode $\psi_k$ satisfies at all times, in momentum
space and in the radiation gauge, the evolution equation:
\beq
\ddot\psi_k + \left[k^2- e^{\phi/2}\left(e^{-\phi/2}
\right)^{\bullet \bullet}\right]
\psi_k=0.
\label{29}
\eeq
In the pre-big bang phase, the general solution of this equation,
normalized to a vacuum
fluctuation spectrum, can be written in terms of Hankel functions
of the second kind as:
\beq
\psi_k= \eta^{1/2} H^{(2)}_\mu (|k\eta|) , \,\,\,\,\,\,\,\,\,\,
\mu=\left |\beta -{1\over 2}\right| , \,\,\,\,\,\,\,\,\,\,
\eta <\eta_1 .
\label{210}
\eeq
In the radiation era we have instead the free
plane-wave solution,
\beq
\psi_k= {1\over \sqrt k}\left[c_+(k) e^{-ik\eta}+
c_-(k) e^{ik\eta}\right] , \,\,\,\,\,\,\,\,\,\,\,\,
\eta >\eta_1 .
\label{211}
\eeq
By matching the two solutions at the transition time $\eta_1$
we easily obtain,
for $|k\eta_1|\ll1$ and $\eta>\eta_1$,
\beq
c_\pm=\pm c(k) e^{\pm ik\eta}, ~~~~~~~~~
\psi_k= {c(k)\over \sqrt k}\sin k|\eta-\eta_1|, ~~ ~~~~~~~
|c(k)|\simeq (k/k_1)^{-\mu-1/2} ,
\label{212}
\eeq
 where $k_1=1/|\eta_1|$ represents the maximal amplified
frequency (higher-frequency modes are unaffected by the
background transition).
The associated energy-density distribution of the produced photons
is then \cite{7}:
\beq
{d\rho(k)\over d\log k}\simeq \left(k\over a\right)^4
|c_-(k)|^2 \simeq \left(k_1\over a\right)^4
\left(k\over k_1\right)^{3-2\mu} , \,\,\,\,\,\,\, k<k_1~,~~~~~\mu<3/2,
\label{213}
\eeq
where $\mu <3/2$ to avoid photon overproduction which would
destroy the homogeneity of the classical background, and where the
amplitude $c(k)$ has been estimated modulo numerical factors of
order 1. At large times $\eta \gg |\eta_1|$ we thus obtain, in string
cosmology,  a cosmic background of electromagnetic fluctuations that,
 for a long enough pre-big bang phase with $\b~ \laq ~2$, are
characterized by a rather flat spectrum, and could provide the
long-sought  origin of the galactic magnetic fields \cite{7}. The amplified
fluctuations satisfy  stochastic correlation functions,  as a
consequence of their quantum origin.

Correspondingly, if we consider axionic perturbations, we are led to
the canonical equation
\beq
\ddot \psi_k+\left(k^2-{\ddot a_A\over
a_A}\right) \psi_k =0,
\label{42a}
\eeq
very similar to Eq. (\ref{29}). The same procedure as in the
electromagnetic case then leads to the spectrum (\ref{213}) with
$\mu=|r|$, where $r$ parametrizes the three-dimensional axion scale
factor as $a_A(\eta)\sim \eta^{r+1/2}$.
For $r=-3/2$, in particular, the axion metric describes a de
Sitter inflationary expansion, and the energy density of a
massless axion background has a flat spectral distribution,
$d\rho/d\log k\simeq (k_1/a)^4$, as first noted in \cite{8}.
The value of $r$ depends on the number and on the kinematics of the
internal dimensions, and the value $-3/2$ can be obtained, in
particular, for a ten-dimensional background 
with special symmetries \cite{B2}.

In the axion case, however, the low frequency tail of the spectrum is
further affected by the radiation $\ra$ matter transition, as the axion
pump field $a_A$ is not a constant (unlike the dilaton) in the
matter-dominated era, where $a_A=a\propto
\eta^2$. This has important consequences that will be discussed in
detail in subsection~\ref{III3}.

After  these preliminary observations  we  shall now
estimate the form of the seed functions for both EM and AX seeds.

\subsection {Electromagnetic seeds}
\label{III2}

Here we determine  the spectral components of the
inhomogeneous stress tensor, for a  stochastic background
obtained by amplifying the quantum EM fluctuations of the
vacuum, as discussed in the previous subsection.
However, independently of the
production mechanism,  the results of this section can be
applied to any EM fluctuation background parametrized by a vector
potential that, in momentum space and in the radiation
gauge, takes the form
\beq
A_i (\bk, \eta) = {c_i (\bk)\over \sqrt k} \sin k\eta ,
\,\,\,\,\,\,\,\,\, k_iA_i=0, \,\,\,\,\,\,\,\,\, A_0=0 \,.
\label{21}
\eeq
$A_i$ is a Gaussian random variable which obeys the stochastic
average condition:
\beq
\langle A_i(\bk)A_j^\ast(\bk')\rangle= {(2\pi)^3\over 2}
\da^3(k-k')\left(\da_{ij}-{k_ik_j\over k^2}\right)
\left|{\bf A}(\bk, \eta)\right|^2.
\label{22}
\eeq
The above condition has been normalized in such a way that
\beq
\sum_i\langle A_i(\bk)A_i^\ast(\bk')\rangle= {(2\pi)^3}
\da^3(k-k')\left|{\bf A}(\bk, \eta)\right|^2 .
\label{23}
\eeq

Taking into account that the electric component of the stochastic
background is rapidly dissipated, because of the conductivity of the
cosmic plasma \cite{10}, the seed stress tensor can be expressed in
terms of the magnetic field only. Setting
$B_i(k)= i\ep_{ijl}k_jA_l(k)$, the
condition (\ref{22}) implies
\beq
\langle B_i(\bk)B_j^\ast(\bk')\rangle= {(2\pi)^3\over 2}
\da^3(k-k')\left(\da_{ij}-{k_ik_j\over k^2}\right)
\left|{\bf B}(\bk, \eta)\right|^2 ,
\label{24}
\eeq
where
\beq
\left|{\bf B}(\bk, \eta)\right|^2=
k^2\left|{\bf A}(\bk, \eta)\right|^2=
k\left|{\bf c}(\bk)\right|^2 \sin^2 k\eta .
\label{25}
\eeq
In a process of photon production, the coefficient
$\left|{\bf c}(\bk)\right|^2$
represents the Bogoliubov coefficient \cite{22b} fixing the  average
photon number density, $\langle n(k) \rangle $, and is linked to the
spectral energy distribution by
\beq
{d\rho(k)\over d\log k}= \left(k\over a\right)^4
{\langle n(k) \rangle\over \pi^2} \simeq
\left(k\over a\right)^4
{\left|{\bf c}(\bk)\right|^2\over \pi^2}.
\label{26}
\eeq
In what follows we shall use for $\left|{\bf c}(\bk)\right|^2$ a
power-law spectrum, characterized by a cut-off frequency
$k_1$,
\beq
\left|{\bf c}(\bk)\right|^2=\left\{\begin{array}{ll}
	 \left(k/k_1\right)^{-2\mu-1}, ~,& k\le k_1, ~~\mu\le 3/2\\
	0 ~,& k> k_1 ~.\end{array} \right.
\label{28}
\eeq
This reproduces in
particular the spectral distribution (\ref{213}), where $\mu$ is fixed
by the dilaton growth rate.

We shall now compute the two-point correlation functions, for
the various components of the inhomogeneous stress tensor
$T_\mu^\nu$, associated with the electromagnetic background:
\beq
\xi_\mu^\nu(x,x')=\langle T_\mu^\nu (x)  T_\mu^\nu (x')
\rangle- \langle T_\mu^\nu (x)\rangle \langle  T_\mu^\nu (x')
\rangle
\label{214}
\eeq
(no sum over $\mu, \nu $, and the angular brackets denote stochastic
average). The Fourier transform of $\xi$ is related to the scalar
 seed functions
$f_\r, f_v, f_\pi$, defined in the previous section.
For $\xi_0^0$ we  have, for instance,
\beq
\xi_0^0(x,x')= \left(M\over a\right)^4\int {d^3k\over
(2\pi)^3}e^{i\bk \cdot ({\bf x}-{\bf x}')} |f_{\rho} (k)|^2 .
\label{215}
\eeq

For $E_i=0$, in particular,  we have to compute  the
correlation of a sum of terms that are quadratic in the
 magnetic field.  We start considering the energy-density  correlation
function,
\beq
\xi_0^0(x,x')=\langle \rho (x)  \rho (x')
\rangle-\left(\langle \rho (x)\rangle \right)^2, \,\,\,\,\,\,\,\,\,
\rho= -T_0^0={|{\bf B}|^2 \over 8\pi a^4} ,
\label{216}
\eeq
and compute
\beq
\Delta^B_{ij}(x,x')=\langle B^2_i (x)  B_j^2 (x')
\rangle-\langle B_i^2\rangle  \langle B_j^2\rangle
\label{217}
\eeq
where, using the stochastic average (\ref{24}) and the reality
condition $B^\ast (k)= B(-k)$,
\beq
\langle B_i^2(x)\rangle  ={1\over 2}\int {d^3k\over (2\pi)^3}
\left| {\bf B}(k)\right|^2 \left(1-{k_i^2\over k^2}\right) .
\label{218}
\eeq
In momentum space, the two-point correlation function for the
energy density can be written as a four-point correlation
function for the stochastic fields (see also \cite{9}). We have, in
particular, \beq
\langle B^2_i (x)  B_j^2 (x')\rangle =\int {d^3k\over (2\pi)^3}
{d^3k'\over (2\pi)^3}{d^3p\over (2\pi)^3}{d^3q\over (2\pi)^3}
e^{i(\bk \cdot
{\bf x}+ \bk' \cdot {\bf x}')}
\langle B_i (p)  B_i (k-p) B_j(q)  B_j (k'-q)\rangle .
\label{219}
\eeq
Decomposing the four-point bracket of the Gaussian variables $B_j$ as
\bea
&&\langle B_i (p)  B_i (k-p)\rangle\langle B_j(q)  B_j
(k'-q)\rangle+ \nonumber \\
&&+\langle B_i (p)  B_j (q)\rangle\langle B_i(k-p)B_j
(k'-q)\rangle +
\langle B_i (p)  B_j(k'-q)\rangle\langle B_i(k-p)
B_j (q)\rangle ,
\label{220}
\eea
and using Eq. (\ref{24}), we find that the first term in the above
equation is exactly cancelled by the quadratic averages
$\langle B_i^2\rangle  \langle B_j^2\rangle$, while the other two
terms give (no sum over $i,j$):
\beq
\Delta^B_{ij}(x,x')=
{1\over 2}\int {d^3k\over (2\pi)^3}{d^3p\over (2\pi)^3}
e^{i\bk \cdot \Delta {\bf x}}
\left| {\bf B}({\bf p})\right|^2 \left| {\bf B}(\bk-{\bf p})\right|^2
\left(\da_{ij}-{p_ip_j\over p^2}\right)
\left(\da_{ij}-{(k-p)_i(k-p)_j\over |\bk-{\bf p}|^2}\right) ,
\label{221}
\eeq
where $\Delta x=x-x'$. By summing over the vector components
we obtain:
\bea
&&\Delta^B(x,x')=\sum_{ij}\Delta^B_{ij}(x,x')=
\nonumber\\
&&={1\over 2}\int {d^3k\over (2\pi)^3}{d^3p\over (2\pi)^3}
e^{i\bk \cdot \Delta {\bf x}}
\left| {\bf B}({\bf p})\right|^2\left| {\bf B}(\bk-{\bf p})\right|^2
\left[1+{|{\bf p} \cdot (\bk -{\bf p})|^2\over
p^2 |\bk-{\bf p}|^2} \right]~.
\label{223}
\eea
According to Eq.~(\ref{215}),  the energy-density
spectrum of the electromagnetic seeds is thus determined by
\beq
\left| f_\rho\right|^2 \left(M\over a\right)^4
={1\over 2 (8\pi a^4)^2}\int
{d^3p\over (2\pi)^3}
\left| {\bf c}({\bf p})\right|^2\left| {\bf c}(\bk-{\bf p})\right|^2  
p|\bk-{\bf p}|\left(1+\cos^2\a\right)
\sin^2p\eta \sin^2|\bk-{\bf p}|\eta ~,
\label{225}
\eeq
where $\a$ is the angle between ${\bf p}$ and $\bk-{\bf p}$.
Inserting the power spectrum (\ref{28}), and defining ${\bf y}=
{\bf p}/k_1$, ${\bf z}=\bk/k_1$,
the above integral can be written, in polar coordinates, as
\beq
\left| f_\rho\right|^2 \left(M\over a\right)^4
={k_1^5\over 2 (8\pi a^4)^2
(2\pi)^2} \int_0^1dy y^{2-2\mu}\int_{-1}^1dx \b^{-2\mu}
\left(1+\cos^2\a\right)\sin^2(yk_1\eta) \sin^2(\b k_1\eta) ,
\label{226}
\eeq
where we defined $x=\cos\vartheta$, $\vartheta$ being  the
angle between ${\bf p}$ and $\bk$, and
\beq
\b^2=|{\bf z}-{\bf y}|^2=y^2+z^2-2xyz , \,\,\,\,\,\,\,\,
\cos^2 \a= \b^{-2}(y^2+x^2z^2-2xyz) ~.
\label{227}
\eeq

The integral of Eq.~(\ref{226}) will be evaluated for
$|k\eta|=|zk_1\eta| \ll 1$, since
we are interested in the large-scale sector of the CMB
anisotropy, namely in the spectrum of all modes that are still
outside the horizon at the time of decoupling.
For EM seeds these modes give the dominant contribution to the
SW effect, as we will see in Section \ref{IV}.
Estimating the
contributions to the
integral from the regions $p\eta \ll 1$, $p\eta \sim 1$ and
$p\eta \gg 1$ , and
recalling that $\mu ~\le ~ 3/2$ according to Eq. (\ref{213}), we find
that the dominant contribution comes from $p\eta \gg 1$
if $\mu\le 3/4$. If $3/4 \le \mu \le 3/2$, the integral is dominated
from its contribution at
 $p\sim k$, thus $p\eta <1$ on super-horizon scales.
 In both cases we
obtain for $f_\r$ a white noise spectrum, i.e. $|f_\r(k)|^2 \sim$
constant, but in the second case there is a
parametric enhancement (see Appendix B). More precisely
\beq
k^3\left| f_\rho\right|^2 \left(M\over a\right)^4
\simeq\left\{ \begin{array}{ll}
 d^2_\rho (k_1/a)^8(k/k_1)^3 , &  \mu\le 3/4 \\
 c^2_\rho (k_1/a)^8(k/k_1)^3 (k_1\eta)^{4\mu-3}, ~~~~
	& 3/4\le \mu\le 3/2 ~,
	\end{array} \right.
\label{228}
\eeq
where  $d_\rho$ and $c_\rho$ are dimensionless numbers of order 1.
Consequently, the energy-density contribution of the EM seeds to the
Bardeen potentials is, according to Eq. (\ref{comp}),
\bea
&&
\ep \eta^2 \left|f_\r\right|k^{3/2}
\simeq\left\{ \begin{array}{ll}
 4 \pi G d_\rho (a\eta)^2(k_1/a)^4(k/k_1)^{3/2} , &  \mu\le 3/4 , \\
  4 \pi G c_\rho (a\eta)^2(k_1/a)^4(k/k_1)^{3/2}
 (k_1\eta)^{2\mu-3/2}, ~~~~
	& 3/4\le \mu\le 3/2 .
	\end{array} \right.
\label{1}
\eea

The contribution of the off-diagonal scalar potential $f_\pi$ can be
similarly obtained by computing the correlation function
$\xi_i^j(x,x')$,
with $i\not= j$.  For purely magnetic seeds,
$f_v=0$, we find
\beq
f_\r=3f_p\sim k^2f_\pi,
\eeq
so that the Bardeen potentials, according to Eq. (\ref{comp}),  are
always dominated by $f_\pi$ on super-horizon scales, as
$\eta^2f_\r/f_\pi \sim (k\eta)^2 \ll1$. Therefore
\bea
&&
k^{3/2}\left|\Psi- \Phi\right| \sim
\ep k^{3/2} \left|f_\pi\right| \simeq \nonumber\\
&&
\simeq \left\{ \begin{array}{ll}
 4 \pi G d_\pi (a\eta)^2(k_1/a)^4(k/k_1)^{-1/2}
(k_1\eta)^{-2} , &  \mu\le 3/4 ,\\
  4 \pi G c_\pi (a\eta)^2(k_1/a)^4(k/k_1)^{-1/2}
 (k_1\eta)^{2\mu-7/2}, ~ ~~~~
	& 3/4\le \mu\le 3/2 .
	\end{array} \right.  ~,
\label{3}
\eea
where  $d_\pi$ and $c_\pi$ are dimensionless numbers of order 1.
By assuming that the universe becomes immediately
radiation-dominated at the physical cut-off scale
$H_1=k_1/a_1$, such a fluctuation spectrum can be expressed
in terms of $\Omega_\gamma(\eta)=
(H_1/H)^2(a_1/a)^4$, i.e. of the fraction of critical energy density in
radiation at a given time $\eta$, and of $g_1=H_1/M_p$,
the transition scale in units of the Planck mass $M_p$. Denoting with  
$\om=k/a$ the proper frequency, and using  $\rho_c=3M_p^2 H^2/8\pi$
for the critical density,  we obtain
\beq
k^{3/2}\left|\Psi- \Phi\right|
\sim \left\{ \begin{array}{ll}
g_1^2\Om_\ga(\eta)(\om/\om_1)^{-1/2}
(\om_1/H)^{-2} , &  \mu\le 3/4 ,\\
g_1^2\Om_\ga(\eta)(\om/\om_1)^{-1/2}
(\om_1/H)^{2\mu-7/2}, ~ ~~~~
	& 3/4\le \mu\le 3/2 .
	\end{array} \right.
\label{229}
\eeq

\subsection {Axionic seeds}
\label{III3}

As a second example of seed inhomogeneities we will consider a
pseudoscalar stochastic background, amplified according to the
perturbation equation (\ref{42a}).

In the initial, higher-dimensional pre-big bang phase, i.e. for
$\eta<\eta_1$, the solution for the canonical variable $\psi$ can be
written as in Eq. (\ref{210}), with $\mu=|r|\leq 3/2$, as discussed
previously.  In the radiation era, i.e. for $\eta_1<\eta<\eta_{eq}$, the
effective potential $\ddot a_A/a_A$ is vanishing, as $\phi=$const and
$a\sim \eta$, and $\psi$ is given by the plane-wave solution
(\ref{211}). In the final matter-dominated era, i.e. for
$\eta>\eta_{eq}$, we have $a\sim \eta^2$, and $\ddot
a_A/a_A=2/\eta^2$. The plane-wave solution is still valid for modes
with $k>k_{eq}=\eta_{eq}^{-1}$, that are unaffected by the last
transitions. Modes with $k<k_{eq}$ feel instead the effect of the
potential in the matter era, and the general solution of Eq. 
(\ref{42a}), for those modes, can be written as
\bea
\psi_k(\eta)&=&{\sqrt{k\eta}\over \sqrt k}\left(AH_{3/2}^{(2)}+B
H_{3/2}^{(1)}\right)\nonumber\\
&=&{\sqrt{k\eta}\over \sqrt k}\left[(A+B)J_{3/2}-i(A-B)Y_{3/2}\right],
~~~~~~k<k_{eq}, ~~~~\eta>\eta_{eq}.
\label{337}
\eea
Here $J_{3/2}$ and $Y_{3/2}$ are Bessel functions of argument $k\eta$
(we follow the conventions of \cite{11}).

The matching of the solutions at $\eta_1$ determines the coefficients
$c_\pm(k)$ as in eq. (\ref{212}). The matching at $\eta_{eq}$ gives
\beq
A+B \sim c({\bf k})\left(k\eta_{eq}\right)^{-1}, ~~~~~~~~~
A-B \sim c({\bf k})\left(k\eta_{eq}\right)^{2},
\label{338}
\eeq
so that the contribution of $J_{3/2}$  to  $\psi_k$ is always dominant
with respect to the $Y_{3/2}$ contribution, both for $k\eta>1$ and
$k\eta<1$. In the matter-dominated era, i.e. for $\eta>\eta_{eq}$, we
can thus approximate the produced stochastic axion background as
follows:
\bea
\sg (\bk, \eta) &\simeq& {c (\bk)\over a\sqrt k} \sin k\eta ,
~~~~~~~~~~~~~~~~k>k_{eq}, \nonumber\\
&\simeq&{c (\bk)\over a\sqrt k}\left(k\over k_{eq}\right)^{-1}(k\eta)^2,
~~~~~k<k_{eq}, ~~~k\eta<1, \nonumber\\
&\simeq&{c (\bk)\over a\sqrt k}\left(k\over k_{eq}\right)^{-1},
~~~~~~~~~~~~k<k_{eq}, ~~~k\eta>1.
\label{339}
\eea

The correlation functions for the various components of the stress
tensor,
\beq
T_\mu^\nu=\pa_\mu\sg\pa^\nu\sg-{1\over 2}\da_\mu^\nu
\left(\pa_\a \sg\right)^2
\label{340}
\eeq
 can be computed by exploiting the stochastic average conditions of
the Gaussian variables $\sg, \dot \sg$ and $\sg_j=\pa_j \sg$,
\bea
&&
\langle \sg(\bk)\sg^\ast(\bk')\rangle= {(2\pi)^3}
\da^3(k-k')\Sg_1(\bk, \eta) ,
\nonumber\\
&&\langle \dot\sg (\bk){\dot \sg }^{\ast} (\bk')\rangle= {(2\pi)^3}
\da^3(k-k')
\Sg_2(\bk, \eta) ,
\nonumber \\
&&\langle \sg_i(\bk)\sg_j^\ast(\bk')\rangle= k_ik_j{(2\pi)^3}
\da^3(k-k')
\Sg_1(\bk, \eta) ,
\nonumber \\
&&\langle \sg_j(\bk){\dot\sg }^\ast(\bk')\rangle=
-\langle \dot\sg (\bk)\sg_j^\ast(\bk')\rangle=
{i} k_j{(2\pi)^3}\da^3(k-k')
\Sg_3(\bk, \eta)~,
\label{341}
\eea
where, according to Eq. (\ref{339}),
\bea
\Sg_1(\bk, \eta)&\simeq&
{\left|{c}(\bk)\right|^2 \over k a^2},
~~~~~~~~~~~~~~~~~~~~~~k>k_{eq}, \nonumber\\
&\simeq&{\left|{c}(\bk)\right|^2 \over k a^2}
\left(k\over k_{eq}\right)^{-2}(k\eta)^4,
~~~~~k<k_{eq}, ~~~k\eta<1, \nonumber\\
&\simeq&{\left|{c}(\bk)\right|^2 \over k a^2}
\left(k\over k_{eq}\right)^{-2},
~~~~~~~~~~~~k<k_{eq}, ~~~k\eta>1
\label{342}
\eea
\bea
\Sg_2(\bk, \eta)&\simeq&
{k}{\left|{c}(\bk)\right|^2\over a^2},
~~~~~~~~~~~~~~~~~~~~~~~~k>k_{eq}, \nonumber\\
&\simeq&0, ~~~~~~~~~~~~~~~~~~~~~~~~~~~~~
~~~~~k<k_{eq}, ~~~k\eta<1, \nonumber\\
&\simeq&{k}{\left|{c}(\bk)\right|^2\over a^2}
\left(k\over k_{eq}\right)^{-2},
~~~~~~~~~~~~k<k_{eq}, ~~~k\eta>1,
\label{343}
\eea
\bea
\Sg_3(\bk, \eta)&\simeq&
{\left|{c}(\bk)\right|^2\over a^2},
~~~~~~~~~~~~~~~~~~~~~~k>k_{eq}, \nonumber\\
&\simeq&0, ~~~~~~~~~~~~~~~~~~~~~~~~~~~~~
~~~k<k_{eq}, ~~~k\eta<1, \nonumber\\
&\simeq&{\left|{c}(\bk)\right|^2\over a^2}
\left(k\over k_{eq}\right)^{-2},
~~~~~~~~~~~~k<k_{eq}, ~~~k\eta>1,
\label{344}
\eea
Following the same procedure as the one used
for EM seeds, and collecting all  contributions to the correlation
function of the axion energy density,
\beq
\rho_\sg={1\over 2 a^2}\left[\dot{\sg }^2 +({\pa}_i
\sg)^2\right] ,
\label{345}
\eeq
we  obtain from $\xi_0^0(x,x')$ that the energy density spectrum is
determined by
\bea
k^3|f_\r|^2\left(M\over a\right)^4&=&{2k^3\over (2a^2)^2}
\int{d^3p\over (2\pi)^3}\Bigg[
\Sg_2({\bf p})\Sg_2({\bf k}-{\bf p})
+\left|{\bf p}\cdot (\bk-{\bf p})\right|^2
\Sg_1({\bf p})\Sg_1({\bf k}-{\bf p})
\nonumber\\
\qquad &-&
2 {\bf p}\cdot (\bk-{\bf p})
\Sg_3({\bf p})\Sg_3({\bf k}-{\bf p})\Bigg].
\label{346}
\eea

In order to evaluate this integral outside the horizon, in the region  
 $k\eta
\leq 1$, we must distinguish two cases, $\mu<3/4$ and $\mu>3/4$. In
both cases, by separate integration in the  ranges  $0<p<\eta^{-1}$,
$\eta^{-1}<p<k_{eq}$, $k_{eq}<p<k_1$, we find a white noise
spectrum,$|f_\rho| \sim$ const. In particular (see Appendix B):
\beq
k^{3/2}\left| f_\rho\right| \left(M\over a\right)^2
=\left\{ \begin{array}{ll}
 d^\sg_\rho (k_1/a)^4(k/k_1)^{3/2}\left[1+\da_\r^\sg
(k_{eq}/k_1)^2(k_1\eta)^{2\mu+1/2}\right], &
~~~  \mu\le 3/4 \\
 c^\sg_\rho (k_1/a)^4(k/k_1)^{3/2}
(k_{eq}/k_1)^2(k_1\eta)^{2\mu+1/2},
	& 3/4\le \mu\le 3/2 ,
	\end{array} \right.
\label{347}
\eeq
where  $c_\rho^\sg, d_\rho^\sg, \da_\r^\sg$ are dimensionless
numbers of order $1$. The same power spectrum  is also
obtained for the scalar velocity potential $f_v$, associated to
the axion seeds. An explicit computation
gives in fact $k f_v \sim k\eta f_\r$ so that
the contribution of $f_\r$ and $f_v$ to the Bardeen potential are both
of the same order, namely:
\bea
&&
\ep \eta^2 \left|f_\r\right|k^{3/2}
\sim \ep \eta^2 {\dot a\over a}\left|f_v\right|k^{3/2}
 =\nonumber\\
&&
=\left\{ \begin{array}{ll}
 4 \pi G d_\rho^\sg (a\eta)^2(k_1/a)^4
(k/k_1)^{3/2}\left[1+\da_\r^\sg
(k_{eq}/k_1)^2(k_1\eta)^{2\mu+1/2}\right], ~~~~~~~& \mu\le 3/4 \\
  4 \pi G c_\rho^\sg (a\eta)^2(k_1/a)^4(k/k_1)^{3/2}
(k_{eq}/k_1)^2(k_1\eta)^{2\mu+1/2},
	& 3/4\le \mu\le 3/2.
	\end{array} \right.
\label{348}
\eea

We will now consider the
anisotropic stress potential $f_\pi$, defined according to
(2.23) by:
\beq
\nabla^4f_\pi={3\over
2 M^2}\pa_i\pa_j\left[\sg_i\sg_j-{1\over
3}\da_{ij}(\pa_k\sg)^2\right] ,  ~~~~\nabla^2= \da_{ij} \pa_i\pa_j.
\label{349}
 \eeq
Summing all contributions in the two point correlation function, we
find
\bea
&&
\langle \nabla^4f_\pi (x) \nabla^4f_\pi (x')\rangle-
\left(\langle\nabla^4f_\pi\rangle\right)^2=
{9\over 2 M^4}\int {d^3k\over (2\pi)^3}
e^{i\bk \cdot \Delta {\bf x}}k^4 \nonumber\\
&&
\int{d^3p\over (2\pi)^3}p^2
|\bk-{\bf p}|^2 \Sg_1({\bf p})\Sg_1({\bf k}-{\bf  
p})\left(\cos^2\vartheta \cos^2\ga -{1\over 3}\cos\vartheta
\cos \ga \cos \a +{1\over 9} \cos^2\a \right),
\label{350}
\eea
where $\vartheta$, $\a$ and $\ga$
are, respectively, the angles between $\bf p$ and $\bk$,
${\bf p}$ and $\bk-{\bf p}$ and $\bk$ and $\bk-{\bf p}$.
The integral over $p$ is of the same type as the integral for the energy
density spectrum (see Eq. (\ref{346})), and gives for $k^2f_\pi$ the
same white noise spectrum (\ref{347}) as for $f_\r$  (modulo numbers
of order $1$) , since
\beq
k^{3/2}\left|f_\pi (k)\right| \left(k\over a\right)^2 M^2\sim
k^{3/2}\left|f_\r(k)\right|\left(M\over a\right)^2.
\label{351}
\eeq
On super-horizon scales the contribution of $f_\pi$ to the Bardeen
potentials is always
dominant with respect to the $f_\rho$ contribution since, from the
above equation,
\beq
 \eta^2f_\rho \sim (k\eta)^2f_\pi.
\label{352}
\eeq
In the whole range $k\eta \leq 1$ we can thus estimate the 
scalar perturbation spectrum, induced by massless axion
seeds, through the $f_\pi$ contribution to the
Bardeen potentials. We find
\bea
&&
k^{3/2}\left|\Psi- \Phi\right| \sim
\ep k^{3/2} \left|f_\pi\right| =\nonumber\\
&&
\left\{ \begin{array}{ll}
 4 \pi G d_\pi^\sg (a\eta)^2(k_1/a)^4(k/k_1)^{-1/2}
(k_1\eta)^{-2}\left[1+\da_\r^\sg
(k_{eq}/k_1)^2(k_1\eta)^{2\mu+1/2}\right], & \mu\le 3/4, \\
  4 \pi G c_\pi^\sg (a\eta)^2(k_1/a)^4(k/k_1)^{-1/2}
(k_{eq}/k_1)^2 (k_1\eta)^{2\mu-3/2}, & 3/4\le \mu\le 3/2,
	\end{array} \right.
\label{353}
\eea
\bea
&&
\sim \left\{ \begin{array}{ll}
g_1^2\Om_\ga(\eta)(\om/\om_1)^{-1/2}
(\om_1/H)^{-2}\left[1+\da_\r^\sg
(\om_{eq}/\om_1)^2(\om_1/H)^{2\mu+1/2}\right], & \mu\le 3/4 ,\\
g_1^2\Om_\ga(\eta)(\om/\om_1)^{-1/2}
(\om_{eq}/\om_1)^2(\om_1/H)^{2\mu-3/2}, & 3/4\le \mu\le 3/2 .
	\end{array} \right.
\label{354}
\eea
where $c_\pi^\sg, d_\pi^\sg, \da_\pi^\sg$ are dimensionless numbers
of order $1$. As we will see in Section \ref{IV}, the dominant  
contribution to the SW effect now comes, for each mode, from the time  
of reentry
$\eta\sim 1/k$.

Let us finally discuss the case of massive axions, with
\beq
T_\mu^\nu=\pa_\mu\sg\pa^\nu\sg-{1\over 2}
\da_\mu^\nu \left[
(\pa_\a\sg)^2-m^2\sg^2 \right] ,
\label{412r}
\eeq
and a primordial distribution again characterized by the index $\mu$,
as in Eq.~(\ref{28}). The mass term directly contribute to $f_\r$ and
$f_p$, and only indirectly to the off-diagonal potentials $f_v$, $f_\pi$.
We are interested in the axion perturbations that may be relevant to
the large-scale CMB anisotropy, namely in the modes that are
outside the horizon at the decoupling era, $k <a H_{dec}$. If, for
these  modes, the mass contribution is negligible, $ma <k<a
H_{dec}$, then the AX seed functions and the corresponding Bardeen
potentials
are the same as in the massless case (see before). We will thus
concentrate our discussion on the case in which the axion mass is
large enough, so that all modes outside the horizon at the
equilibrium epoch are already non-relativistic:
\beq
ma > aH_{eq}> k.
\label{413a}
\eeq
In this case we may neglect the effects of an additional axion
production in the matter-dominated era, since
$a^2m^2>\ddot{a}/a$ at $\eta\geq \eta_{eq}$. The axion fluctuations
are amplified by the inflation $\ra$ radiation transition, but are to  
be evaluated in
the non-relativistic regime ($\eta>\eta_{eq}$), where the mass
contribution is already important.

For non-relativistic, super-horizon modes, the Fourier components of
the axion field become (see the non-trivial calculation
reported in  Appendix~C):
\beq
\sg (\bk, \eta) = {c (\bk)\over a\sqrt {ma}}
\left(k\over k_1\right)^{1/2}\left(H_1\over m\right)^{1/4} \sin
\left(m\over H\right) ,~~~~~~k<k_m=k_1\left(m\over
H_1\right)^{1/2},
 \label{414a}
\eeq
where the initial distribution $c(k)$ is still given by  Eq. (\ref{28}).
Here $k_m$ is the limiting frequency re-entering the horizon at the
same time as it becomes non-relativistic, i.e. $k_m/a_m=H_m=m$.
Indeed, we are assuming that at the transition scale $H_1$ the mass
term is completely negligible, $m \ll H_1$, and all modes are
relativistic. As the proper momentum is red-shifted, the modes
become non-relativistic when $m=k/a=\om$, and re-enter the horizon
when $ H=\om$.

For the axion field (\ref{414}) the stochastic conditions (\ref{341}) are
still valid, but the squared amplitude (\ref{342}), averaged over time
scales $m/H\gg 1$, now become
\beq
\Sg_1({\bf p},\eta)= {|c(\bk)|^2\over 2 m a^3}
\left(k\over k_1\right)\left(H_1\over m\right)^{1/2}=
{1\over m^2a^2}\Sg_2({\bf p},\eta)=
{1\over ma}\Sg_3({\bf p},\eta) .
\label{415a}
\eeq

For the case we are considering, the contribution of $f_\pi$ to the
Bardeen potentials is always negligible with respect to $\eta^2 f_\r$.
An explicit computation gives, in fact,
\beq
\eta^2 f_\r/f_\pi \simeq m/H \gg 1,
\label{416}
\eeq
where the last inequality is a consequence of (\ref{413}). In addition,
the mass contribution to the AX energy density dominates with
respect to the momentum contribution, since $ m>k/a$. The energy
density correlation function thus becomes:
\beq
\xi_0^0(x,x')= m^4\left(\langle\sg^2(x)\sg^2(x')\rangle-
\langle \sg^2(x)\rangle^2\right)
\label{417}
\eeq
(as $|\dot \sg(k)|=ma |\sg(k)|$), and gives, using Eq.~(\ref{415}):
\bea
&&
k^3\left| f_\rho\right|^2 \left(M\over a\right)^4
= m^4 k^3
\int{d^3p\over (2\pi)^3}
\Sg_1({\bf p})\Sg_1({\bf k}-{\bf p})
\nonumber\\
&&={mH_1\over 8\pi^2}\left(k_1\over a\right)^6\left(k\over
k_1\right)^3 \int_0^1d y y^{2-2\mu} \int _{-1}^1dx
\b^{-2\mu}
\label{418}
\eea
where $x,y$ and $\b$ are defined in Section \ref{III2}.

It should be noted that the above expression for the spectrum is only
valid if $\mu>3/4$. Only in that case, in fact, is the integral over  
$y$dominated by the contribution of the lower boundary, $p/k_1\ra 0$,
and is the use of Eq.~(\ref{415a}) for the axion spectrum appropriate.
In the opposite case, we have to take into account the different
spectrum of non-relativistic sub-horizon modes, for $p>k_m$, and
possibly of relativistic modes in the high-frequency limit $p\ra k_1$.
In both cases we obtain, for $\mu<3/4$, a white noise spectrum and
a negligible contribution to the large-scale anisotropy, as we will see
in the next section.

We will thus concentrate on the case $3/4<\mu \leq 3/2$. For this case
the integral (\ref{418}) is estimated in Appendix B,
and we obtain
\beq
k^3\left| f_\rho\right|^2 \left(M\over a\right)^4
 = c_m^2mH_1 \left(k_1\over a\right)^6
\left(k\over k_1\right)^{6-4\mu}  , ~~~~~~~3/4<\mu\leq 3/2 \, ,
\label{419}
\eeq
where $c_m$ is  a dimensionless number of order 1. The
corresponding Bardeen spectrum is:
\bea
&&
k^{3/2}\left| \Psi\right| \sim k^{3/2}\left| \Phi\right|\sim
\ep\eta^2  \left|f_\r\right|k^{3/2} =\nonumber\\
&&
= 4\pi G c_m  (a\eta)^2 \left(mH_1\right)^{1/2}
\left(k_1\over a\right)^3
\left(k\over k_1\right)^{3-2\mu} \nonumber \\
&&
\sim g_1^2 c_m\left(m\over
H_1\right)^{1/2}\left(H_1\over H\right)^{2}\left(a_1\over a
\right)^{3}\left(\om\over \om_1\right)^{3-2\mu}\sim \Om_\sg(\om).
\label{420}
\eea
We may note that $\Psi H^2$ evolves in time like $a^{-3}$, so that,  
during the
matter-dominated era (when $H^2\propto a^{-3}$), the Bardeen
potential $\Psi$ remains frozen at the value  reached at the time
$\eta_{eq}$ of matter-radiation equilibrium.  Using
$(H_1/H_{eq})^2(a_1/a_{eq})^3= (a_{eq}/a_1) =
(H_1/H_{eq})^{1/2}$, we obtain for $\eta >\eta_{eq}$,
\beq
k^{3/2}\left| \Psi\right| \sim k^{3/2}\left| \Phi\right|\sim
c_m g_1^2 \left(m\over H_{eq}\right)^{1/2}
\left(\om\over \om_1\right)^{3-2\mu}.
\label{421}
\eeq

The CMB anisotropy induced by the EM
and AX seeds discussed here will be analysed in the next section.

\section {CMB fluctuations from pre-big bang seeds}
\label{IV}

For electromagnetic
seeds, with the assumption that the electric field is already
dissipated away at recombination, we find that
the seeds are generically suppressed by a factor
$(k \eta_{dec})^2$, and the anisotropic stress $f_{\pi}$
dominates over
the density contribution $f_{\rho}$ (see the discussion at the end of
Section \ref{II1}).
By contrast, for massless axionic perturbations,
there is no $ (k \eta_{dec})^2$ suppression for $f_{\rho}$, while
there is one for $f_{\pi}$. For large wave numbers which enter the
horizon before matter and radiation equality, EM and AX seeds lead to
similar amplitudes. Consequently, if the convolution leading to
$f_\pi$ is dominated by  small scale contributions, $\mu<3/4$, the
two cases give similar  geometric scalar perturbations $\Psi, \Phi$,
through Eq.~(\ref{comp}).

However, on large scales, $k\eta_{eq}<1$, the additional axion
production during the matter-dominated era leads to an enhancement
by the factor  $(\eta/\eta_{eq})^2$. This changes the time-dependence
of the Bardeen potentials and has important consequences as we
will see below.

\subsection{Electromagnetic seeds}
\label{IV1}

The scalar metric perturbation spectrum induced by EM seeds is
reproduced in Eqs.~(\ref{3}) and (\ref{229}). Comparing with our
parametrization in terms of $\a$ and $\ga$ (see Eqs. (\ref{252}),
(\ref{253})) we find
\bea
\gamma &=&\left\{\begin{array}{ll}
	 -4, &  \mu\le 3/4\\
	2\mu-11/2, ~~~~~~  &  3/4\le \mu\le 3/2 ~ \end{array} \right.\\
\alpha &=&\left\{\begin{array}{ll}
	 7/2, &  \mu\le 3/4\\
	5-2\mu , ~~~~~~ & 3/4\le \mu\le 3/2 ~ \end{array}\right.
\eea
and \beq
{\cal N} =  c_\pi\left(g_1\over4\pi\right)^2(k_1\eta_{eq})^2
\eeq
in both cases $\mu <3/4, \mu>3/4$ (modulo numbers of order $1$).

Since $\ga+1<0$, in both cases the seeds decay
fast enough outside the horizon, and our analysis of Section \ref{II} 
applies. However, in both cases $\ga+\al=-0.5$, which leads to the
spectral index $n=2$, i.e. to a spectrum that grows too fast
with frequency to fit the results of COBE observations,
see Eqs.~(\ref{indexa}), (\ref{273}).

The quadrupole amplitude is given by
$Q_{rms-PS}=\sqrt{(5/4\pi)C_2} T_0$, which has been measured
\cite{banday}
to be $ Q_{rms-PS} = (18\pm 2) \mu K$. This leads to
\beq
C_2 = (1.09  \pm  0.23)\times 10^{-10}  ~. \label{55}
\eeq 
From Eq.~(\ref{C_ell_sw}), using
$\al+\ga=-1/2$, $k_1\eta_{eq}=(H_1/H_{eq})^{1/2}$,
$g_1=H_1/M_p$, and setting $\ell=2$, we obtain:
\beq
C_2^{SW} \approx
{c_\pi^2 g_1^{6-\a}\over
10(4\pi)^4(\ga+1)^2}
\left(M_p\over H_{eq}\right)^{2-\al}
	\left({\eta_{dec}\over\eta_0}\right)^{2(\ga+1)}
	\left({\eta_{eq}\over\eta_0}\right)^{2\al} ~.
\label{56}
\eeq
Compatibility with the COBE normalization, $C_2 ~\laq ~10^{-10}$,
thus implies
\beq
(6-\a)\log_{10} g_1 ~\laq~ 55(\a-2) -6 + \log_{10}(\ga+1)^2 -
\log_{10}c_\pi^2
\label{57}
\eeq
(we have used $H_{eq}/M_p \sim 10^{-55}$, and $\eta_{dec}\sim
\eta_{eq}\sim 10^{-2}\eta_0$).  This important constraint is easily
satisfied by a growing seed spectrum, $\mu<3/2$, i.e. $\a>2$. In the
limiting (and most unfavorable) case $\mu=3/2$, \, $\a=2$, \,
$\ga=-5/2$, the above condition reduces to
\beq
\log_{10} g_1 ~\laq~ -1.4 -0.5
\log_{10}c_\pi ~.
\label{58}
\eeq
Even in this limiting case there are no stringent constraints on the
typical inflation scale of the ``minimal" pre-big bang scenario
\cite{1,5b,4a},
expected to approach the string mass scale $M_s$ as $g_1=H_1/M_p
\sim M_s/M_p$. Indeed, the limiting condition (\ref{58}) is
marginally compatible even with the maximal expected value $H_1\sim
M_s$, since \cite{30a}
\beq
10^{-2} ~\laq~ {M_s/M_p}~ \laq~ 10^{-1}.
\label{59}
\eeq

To conclude, the EM fluctuations seem to lead to a scalar
perturbation spectrum that grows too fast with frequency to
contribute in a significant way to the observed large-scale
anisotropy. The positive aspect of our result is that 
there are no significant constraints from
the COBE normalization to the production of seeds for galactic
magnetic fields, which remains allowed as discussed in \cite{7}.

\subsection{Axionic seeds}
\label{IV2}

Let us first consider massless axions.
If $\mu < 3/4$, the situation is like in the electromagnetic
case. The CMB fluctuations induced have the wrong spectrum, but their
amplitude is sufficiently low to avoid conflict with
observations.

If $3/4\leq\mu\leq 3/2$ the situation becomes radically different.
Comparing Eq.~(\ref{353}) with the ansatz (\ref{252}),~(\ref{253})
we obtain, due to the additional factor $(\eta/\eta_{eq})^2$,
\bea
	\ga &=& 2\mu-7/2, \\
	\a &=& -\ga - 1/2 = -2\mu +3  ~.
\eea
In the limiting case $\mu=3/2$ this yields $\ga=-1/2$ and $\al=0$,
which corresponds to a Harrison-Zel'dovich spectrum of CMB
fluctuations, according to Eq. (\ref{index>-1}), with an amplitude
\beq
 {\cal N} \simeq g_1^2
\label{411}
\eeq
(we have absorbed into $g_1$ all dimensionless numerical coefficients
of order one appearing in the spectrum (\ref{353})). Note that
$f_\r$ leads to a Bardeen potential  with the same $\a$, but with
$\ga=2\mu-3/2$. However, since again $\ga>-1$, the contribution to  
the SW  effect is the same for $f_\r$ and $f_\pi$
(see Section \ref{II}).

The normalization of the axion spectrum to the COBE amplitude
(\ref{55}), according to Eq. (\ref{simple}), imposes the condition
\beq
C_2^{SW}\simeq  {\cal N}^2 \left(k_1\eta_0\right)^{-2\a}\simeq
g_1^4\left(\om_0\over \om_1\right)^{6-4\mu}\simeq 10^{-10},
\label{412}
\eeq
which implies
\beq
\log_{10}g_1\simeq {164-116 \mu\over 1+2\mu}
\label{413}
\eeq
(again we have absorbed numerical coefficients into $g_1$, and we
have used $\om_0\sim 10^{-18}$ Hz, $\om_1 \sim g_1^{1/2} 10^{11}$
Hz, according to \cite{5b,4a}). On the other hand, the allowed range
for the spectral index (see Eq. (\ref{index>-1}), combined with the
condition $\mu\leq 3/2$ (required to prevent over-critical axion
production), implies
\beq
	1.4<\mu<1.5 \label{414}.
\eeq
The combination of (\ref{413}), (\ref{414}) leads to
\beq
3\times 10^{-3} ~\laq~g_1=(H_1/M_p)~\laq~2.6,
\label{415}
\eeq
which is perfectly compatible with the identification $H_1\sim M_s$
(see Eq, (\ref{59}).

A stochastic background of massless axions, produced in the context of
the pre-big bang scenario, is thus a possible viable candidate
for a consistent explanation of the large-scale anisotropy observed by
COBE. The important difference between AX and EM seeds is the
non-conformal coupling of the axions to the metric, that leads to an
additional amplification of perturbations after the
matter-radiation equality.

Another interesting case is that of a massive axion background, for
which the $f_\pi$ contribution to the Bardeen potentials is negligible
when the super-horizon modes are already non-relativistic at the time
of decoupling, $m>H_{dec}$. As seen in the previous section, one then
obtains constant Bardeen potentials, with $\ga=0$,
$\a=(n-1)/2=3-2\mu$ and
\beq
{\cal N}= g_1^2 \left(m\over H_{eq}\right)^{1/2}
\label{510}
\eeq
(see Eq.~(\ref{421}), where we have set $c_m=1$). A flat
Harrison-Zel'dovich spectrum is again possible in the limiting case
$\mu=3/2$. The amplitude of perturbations, however, is enhanced by
the factor $(m/H_{eq})^{1/2}$, so that the value of the axion mass has
to be bounded, to avoid conflicting with the COBE normalization
(\ref{55}).

The allowed range for
the spectral index, $0.8\leq n \leq 1.4$, and the condition $\mu<3/2$ to
avoid over-critical axion
production, again imply for the parameter $\mu$ the allowed range
(\ref{414}).
In addition, the present axion energy density is constrained by the
critical density bound, $\Om_\sg (\eta_0)\leq 1$, imposed at
the peak frequency $\om_m$ of non-relativistic modes
(see Appendix C).  Actually, an even stronger condition is required for
the validity of our perturbative approach, which neglects the
back-reaction of the axionic seeds on the expansion of the universe.
Using Eq.~(\ref{c25})
we thus impose the condition
\beq
\Om_\sg (\om_m,\eta_0) \sim g_1^2
\left(H_{1}\over H_{eq}\right)^{1/2} \left(m\over
H_1\right)^{2-\mu} ~\laq~ 0.1, \label{511a}
\eeq
which implies
\beq
(2-\mu)\left[\log_{10}(m/H_{eq})-\log_{10}g_1-55\right]+{5\over 2}
\log_{10}g_1 +{55\over 2} <-1.
\label{511b}
\eeq

In order to find a possible AX mass window compatible with the
COBE data, we now impose the normalization $C_2 \simeq
10^{-10}$ on the massive axion spectrum Eq.~(\ref{421}). From Eq.
(\ref{simple}) we obtain
\beq
C_{2}^{SW} \simeq {\cal N}^2(k_1\eta_0)^{-2\al}\simeq
	g_1^4 \left(m\over H_{eq}\right)
\left(\om_0\over\om_1\right)^{6-4\mu}\simeq 10^{-10},
 \label{420a}
\eeq
from which
\beq
\mu\simeq\left[ 164 - \log_{10}(m/H_{eq})- 4 \log_{10}g_1  \right]/116
\label{513}
\eeq
(we have used $\om_1/\om_0\sim 10^{29}$, neglecting the weak
dependence of $\om_1$ on the transition scale $g_1$).
By eliminating $\mu$ in terms of $m$ and $g_1$, according to the
above equation, the constraints (\ref{414}) and (\ref{511b}), plus
the condition $m>H_{dec}\sim H_{eq}$
(assumed for the validity of the
spectrum (\ref{421})), determine an allowed region in the plane
$(m,H_1)$ as follows:
\beq
\left\{\begin{array}{ll}
	{10^{-10}}\left(M_p/ H_1\right)^4
~\laq~ {m/ H_{eq}} ~\laq~
{10^{1.6}}\left(M_p/ H_1\right)^4, ~~~~~~~~~~m~\gaq~{H_{eq}}, &  \\
\left[68+\log_{10}(m/H_{eq}) +4 \log_{10}g_1\right]
\left[\log_{10}(m/H_{eq})-55-\log_{10}g_1\right] +58\left(55+
5 \log_{10}g_1\right)\laq -1. &  \end{array}
\right.
\label{514}
\eeq

For a typical string cosmology scale, $H_1 \sim M_s \sim (10^{-1} -
10^{-2}) M_p$, we thus obtain the maximal  allowed window:
\beq
10^{-27} \;{\rm eV} ~\laq~ m ~\laq~ 10^{-17} \;{\rm eV} .
\label{515}
\eeq
As illustrated in Fig. 1,
the window is shifted towards higher values of  mass as the final
inflation scale is lowered, and the  seed condition (\ref{511a})
becomes important only at
low inflation scales, $ H_1/M_p ~\laq~ 10^{-7}$. The stringent upper  
limit we obtained for the mass can be traced back to the simplest  
model of background used in this paper, that gives the same slope for  
the axion spectrum at low and high frequency (see eqs. (\ref{c24}),
(\ref{c26})).
It is not excluded that higher values of the mass may become possible  
in a more complicated model of background, giving a steeper high  
frequency spectrum.

\begin{figure}[t]
\begin{center}
\mbox{\epsfig{file=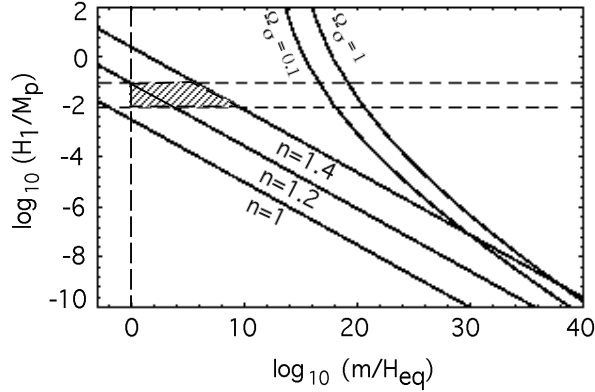,width=82mm}}
\vskip 5mm
\caption{\sl The phenomenologically allowed region is to the left
of the curve $\Om_\sg=0.1$, to the right of the vertical dashed line
$m=H_{eq}$, and lies within the full lines $n=1$, $n=1.4$, to avoid
conflicting with present COBE observations ($n<1$ is excluded by
over-critical axion production). The shaded area defines the  allowed
mass  window for an inflation scale $H_1=M_s$, typical of string
cosmology.}
\end{center}
\end{figure}

\section {Conclusions}
\label{V}

In this paper we have considered the possibility that, in a string
cosmology context, the large-scale temperature anisotropies may
arise from the contribution of seeds to the metric  fluctuations,
and {\em not} from the direct amplification of the metric fluctuations
themselves, as in the conventional inflationary
scenario. We have discussed, in particular, two cases: one in which
the seeds are EM vacuum fluctuations amplified by the growth of the
dilaton field, and one in which the seeds are AX vacuum fluctuations
amplified by the time evolution of a higher-dimensional background.

In the case of EM  perturbations we have found
that the induced angular power spectrum of $\Delta T/T$ grows
too fast to be compatible with COBE observations. However, the
contribution of the seeds to the large-scale anisotropy may be
consistently imposed to be negligible, without constraining in a
significant way the basic parameters of the pre-big bang models.

Massless AX perturbations, unlike EM perturbations, 
are also affected by the radiation $\ra$
matter transitions. This changes the time
dependence of the seed contribution to the Bardeen potentials and, 
due to the integrated Sachs-Wolfe effect, a flat or slightly
tilted blue spectrum of temperature anisotropies can be induced,
compatible with present COBE observations. Scale-invariant massless
axion seeds thus appear as possible promising candidates for structure
formation. Determining in more details the CMB anisotropy spectrum
also on smaller angular scales requires however numerical simulations, 
which we defer to a future research project.

For massive AX seeds the situation is qualitatively different if the 
mass is such that all modes outside the horizon at the time of
decoupling are already non-relativistic. In that case the
contribution to $\Delta T/T$ is controlled by the axion mass, and a
slightly tilted blue spectrum is still compatible with the amplitude and
the slope measured by COBE, provided the axion mass is inside an
appropriate window, in the ultra-light mass region. Higher values of  
masses may become possible in models with  more complicated 
backgrounds. 

At smaller angular scales, an axionic origin of CMB anisotropies
should lead to acoustic  peaks in the  spectrum,
with a structure different  from that of the
standard inflationary scenario. This may in principle allow a test of 
models with axionic seeds through the very accurate
observations of the CMB anisotropy planned in the near future
\cite{30b}. A thorough investigation of this possibility  is postponed to a
future paper.

\acknowledgements

We are grateful to Massimo Giovannini for  helpful discussions.
R. D. and M. S. are partially supported  by the Swiss NSF.
M. S. acknowledges financial support from the Tomalla
Foundation.

\bigskip

\appendix

\section{Sachs-Wolfe coefficients for power-law
spectra}
\label{A}

Assume that the Bardeen potentials are given by
power-law spectra
as in Eq.~(\ref{power-law}),
\beq
\Psi-\Phi =\left\{\begin{array}{ll}
	C(k)x^\ga, & x\ll 1\\
	C(k), & x\gg 1	\end{array} \right. ~,
~~~~~C(k)={\cal N}k^{-3/2} (k/k_1)^{\al} ~,
\eeq
where $x=k\eta, \, x_0=k\eta_0, \, x_{dec}=k\eta_{dec}$.
The SW contribution to the angular coefficients
 $C_\ell$ is given by
\beq
C_\ell^{SW} = {\cal N}^2{2\over \pi}\int_0^{k_1} {dk\over k}
\left({k\over k_1}\right)^{2\al}|I(k)|^2,
\label{intC}
\eeq
where
\beq
I(k)=j_\ell(x_0-1) +
\int_{x_{dec}}^1 x^{\ga} j'_\ell (x_0-x) dx ~,
\eeq
and a prime stands for the derivative of $j_\ell$ with respect to its
argument.

We concentrate here on the  case where $\ga+1<0$. Furthermore,
we are interested in the situation where the integral in
Eq.~(\ref{intC}) is dominated by large scales (small values of $k$),
and therefore $x_{dec} \ll 1$. In that case the integral $I(k)$ is
dominated by its value at the lower bound: 
\beq
I(k) \approx {1\over |1+\ga|}x_{dec}^{\ga+1}j'_{\ell}(x_0) ={1\over
|1+\ga|}x_{dec}^{\ga+1}\left[{\ell\over 2\ell +1}j_{\ell-1}(x_0) -
	{\ell+1\over 2\ell +1}j_{\ell+1}(x_0)\right].
\eeq
This leads to the following expression for the $C_{\ell}$'s:
\bea
C_\ell^{SW} &=& {\cal N}^2{2\over \pi}
	\left({\eta_{dec}\over \eta_0}\right)^{2(\ga+1)}{1\over
|1+\ga|^2}
	(k_1\eta_0)^{-2\al}\int_0^{\infty} {dx_0\over x_0}
	x_0^{2(\al+\ga+1)}\nonumber \\
 && \times\left[{\ell^2\over(2\ell+1)^2}j_{\ell-1}^2(x_0) -
	{2\ell(\ell+1)\over(2\ell+1)^2}j_{\ell-1}(x_0)j_{\ell+1}(x_0) +
	{(\ell+1)^2\over(2\ell+1)^2}j_{\ell+1}^2(x_0)\right] \nonumber\\
 &=&  {{\cal N}^2\over |1+\ga|^2} {2\over \pi}
	\left({\eta_{dec}\over \eta_0}\right)^{2(\ga+1)}
	(k_1\eta_0)^{-2\al}    \nonumber \\
&& \times\left[{\ell^2\over(2\ell+1)^2}I^{(1)}_\ell
     -{2\ell(\ell+1)\over(2\ell+1)^2}I^{(2)}_\ell
     +{(\ell+1)^2\over(2\ell+1)^2}I^{(3)}_\ell
\right] \, ,
\eea
where, setting $j_\ell=\sqrt{\pi/x} J_{\ell-1/2}$, we find (Ref.
\cite{GrRy}, number~6.574) for $\alpha + \ga <0$,
\bea
I^{(1)}_\ell &=& {\pi\over  2}
	\int_0^{\infty}dxx^{2(\al+\ga)}J^2_{\ell-1/2}(x) \nonumber\\
	&=&
	{\pi \over 2} {\Ga(-2(\al+\ga))\Ga(\ell+\al+\ga)\over
	2^{-2(\al+\ga)}[\Ga(-(\al+\ga)+1/2)]^2\Ga(\ell-(\al+\ga))} ~;\\
I^{(2)}_\ell &=& {\pi\over  2}
	\int_0^{\infty}dxx^{2(\al+\ga)}J_{\ell-1/2}(x)J_{\ell+3/2}(x)
	\nonumber\\
	& =&
	{\pi \over 2} {\Ga(-2(\al+\ga))\Ga(\ell+1+\al+\ga)\over
	2^{-2(\al+\ga)}\Ga(-(\al+\ga)-1/2)\Ga(3/2-(\al+\ga))
	\Ga(\ell+1-(\al+\ga))} ~;\\
I^{(3)}_\ell &=& {\pi\over  2}
	\int_0^{\infty}dxx^{2(\al+\ga)}J^2_{\ell+3/2}(x) \nonumber\\
	&=&
	{\pi \over 2} {\Ga(-2(\al+\ga))\Ga(\ell+2+\al+\ga)\over
	2^{-2(\al+\ga)}[\Ga(-(\al+\ga)+1/2)]^2\Ga(\ell+2-(\al+\ga))} ~.
\eea	
 Finally, combining the above results, we
obtain the result given in Eq.~(\ref{C_ell_sw}):
\bea
&&
C_\ell^{SW} = {{\cal N}^2\over 2^{-2(\al+\ga)}(\ga+1)^2}
	{\Ga(-2(\al+\ga))\over\Ga(1/2-\al-\ga)^2}    %%%\nonumber\\  &&
	\left({\eta_{dec}\over \eta_0}\right)^{2(\ga+1)}
	\left(k_1\eta_0\right)^{-2\al}
        {\Ga(\ell+1+\al+\ga)\over \Ga(\ell+1\-\al-\ga)}  \nonumber\\
	&&\times\left[{\ell^2\over (2\ell+1)^2}
          {\ell-\al-\ga\over \ell+\al+\ga}
        +{2\ell(\ell+1)\over (2\ell+1)^2} {1/2+\al+\ga\over
	1/2-\al-\ga} +{(\ell+1)^2\over (2\ell+1)^2}
        {\ell+1+\al+\ga \over \ell+1-\al-\ga}  \right] ~.\label{C_ell_swA}
\eea
It is interesting to note that, for $\al+\ga=-1/2$, the mixed term
$I^{(2)}_\ell$ vanishes, which is indeed what happens in the case of
electromagnetic seeds (see Section \ref{IV}).

\section{The seed functions}
\label{B}
\subsection{Electromagnetic Seeds}
For purely magnetic seeds, all the seed functions can be
approximately determined by the energy density correlation
function $\xi_0^0$, which leads to Eq.~(\ref{226}). The contribution of
super-horizon modes ($k\eta \ll 1$) to the spectrum can be
estimated in the limit $z =k/k_1 \ra 0$. In this limit $\b \ra y$,
$\cos^2\a \ra 1$, and the integral (\ref{226}) reduces to
\beq
I= {k^3 k_1^5\over a^8}\int_0^1 dy~y^{2-4\mu} \sin^4(yk_1\eta),
~~~~~~\mu\leq 3/2.
\eeq
The dominant region of integration is easily shown to
be  $ y \sim 1$ for $\mu\leq 3/4$ and $yk_1\eta \sim 1$ for $3/4 \le
\mu\leq 3/2$. This gives
\beq
I=\left\{\begin{array}{ll}
(k_1/a)^8(k/k_1)^3, &\mu\leq 3/4\\
(k_1/a)^8(k/k_1)^3 (k_1\eta)^{4\mu-3},
 ~~~~& 3/4 \le  \mu\leq 3/2
	\end{array} \right. ~,
\eeq
modulo numerical factors of order one. This coincides with the
result reported in Eq.~(\ref{228}).

\subsection{Massless Axions}

For massless axions, the seed spectral functions are determined by  
the integrals (\ref{346}), (\ref{350}). The various terms appearing in  
the integrands turn out to give comparable contributions, so let us  
concentrate on the typical term
\beq
I={k^3\over a^4}\int d^3p~ p^2 |{\bf k}-{\bf p}|^2 \Sg_1({\bf p})
\Sg_1({\bf k}-{\bf p}).
\eeq
We distinguish different integration regions:
$0<p<k$, $k<p<\eta^{-1}$, $\eta^{-1}<p<k_{eq}$, $k_{eq}<p<k_1$.
The dominant integration regions depend on the value of $\mu$ but,
for all $\mu \leq 3/2$, they always lie at $p \geq \eta^{-1}>k$.
This is the reason why we always obtain a white noise spectrum.
On the other hand, the behaviour in $\eta$  depends on which region
of $p$ dominates. Specifically we find:

1) For $3/4 \leq \mu \leq 3/2$ the leading contribution to $I$
comes from $p\sim \eta^{-1}$, and gives the single term appearing  
in eq. (\ref{347}).

2) For $\mu < 3/4$ the leading contribution comes either from
$p\sim k_1$ (giving the first term in the square brackets of  
(\ref{347})), or (for $\mu$ very close to $3/4$)  
from $p\sim \eta^{-1}$ (giving the second term
in the same brackets).

\subsection{Massive Axions}

For massive actions, the energy density spectrum is determined by Eq.
(\ref{418}), with $3/4<\mu\leq 3/2$. This integral is dominated by the
region $p\sim k$, and its rough behaviour can be easily obtained this
way. For a more precise evaluation we proceed as
follows: the angular integration gives
\beq
k^3 |f_\rho|^2 \left(M\over a\right)^4
= {mH_1\over 16\pi^2 z(\mu-1)}\left(k_1\over a\right)^6\left(k\over
k_1\right)^3 \int_0^1 dy~
y^{1-2\mu} \left[(z-y)^{2-2\mu}-(z+y)^{2-2\mu}\right].
\eeq
Defining $t=y/z$ we obtain
\beq
k^3 |f_\rho|^2 \left(M\over a\right)^4
= {mH_1\over 16\pi^2 (\mu-1)}\left(k_1\over a\right)^6\left(k\over
k_1\right)^3 z^{3-4\mu}
\left(A-B\right)
\eeq
 where, after some manipulation \cite{GrRy},
\bea
A&=&\int_0^\infty dt\,\, t^{1-2\mu}
\left[(1-t)^{2-2\mu}-(1+t)^{2-2\mu}\right] =\nonumber\\
&=&
{2^{4\mu-4}\over \sqrt \pi}\Ga (2-2\mu)\Ga(2\mu-3/4)
\left[\cos 2\pi (\mu-1) -1\right]
\eea
and
\beq
B=\int_{1/z}^\infty dt\,\, t^{1-2\mu}
\left[(1-t)^{2-2\mu}-(1+t)^{2-2\mu}\right] .
\eeq
By evaluating this second integral in the limit $z \ra 0$,  we
obtain
\beq
B\sim z^{4\mu-3} \ll A.
\eeq
so that
\beq
k^3 |f_\rho|^2 \left(M\over a\right)^4
= {mH_1 A\over 16\pi^2 (\mu-1)}\left(k_1\over a\right)^6\left(k\over
k_1\right)^{6-4\mu},
~~~~~~~~~~~~~~~~ 3/4<\mu <3/2 ,
\eeq
as reported in eq. (\ref{419}). Note that there is no singularity for
$\mu=1$, as
\beq
\lim_{\mu \ra 1}{\Ga (2-2\mu)\over (\mu-1)} \left[\cos 2\pi (\mu-1)
-1\right] = {4\pi^2\over (\mu-1)^2}(\mu-1)^2={\rm const}
\eeq

\section{Non-relativistic corrections to the axion spectrum}
\label{C}

For a massive-axion perturbation $\sg$, the string frame action
\beq
S={1\over 2}\int d^4x \sqrt{-g} e^\phi \left[(\pa_\mu \sg)^2
-m^2\sg^2\right],
\eeq
in a conformally flat background, can be written in terms of the
canonical variable
\beq
\psi =z \sg, ~~~~~~~~~~~~~~~~ z= a e^{\phi/2},
\eeq
as
\beq
S={1\over 2}\int d^3x d\eta \left[\dot {\psi}^2 -(\pa_i\psi)^2 +
{\ddot z\over z}\psi^2 -m^2a^2 \psi^2 \right]
\eeq
(the dot denotes differentiation with respect to the conformal time
$\eta$).  The Fourier modes $\psi_k$ satisfy the perturbation
equation
\beq
\ddot \psi_k +\left(k^2 -{\ddot z\over z}+m^2a^2\right)\psi_k=0.
\label{c4}
\eeq

We shall consider the background transition at $\eta=\eta_1$ from
an initial pre-big bang phase in which the axion is massless, to a
final radiation-dominated phase in which the dilaton freezes to its
present value, and the axion acquires a small (in string units) mass.
For $\eta>\eta_1$ the solution of Eq. (\ref{c4}) depends on the
kinematics of the pump field $z$ and, after normalization to an
initial vacuum spectrum, it can be written in terms of the
second-kindHankel functions \cite{11} as:
\beq
\psi_k(\eta)=\eta^{1/2}H_\mu^{(2)} (k\eta)
\label{c5} .
\eeq
In the radiation era, $\eta>\eta_1$, the ``effective potential"
${\ddot z/z}$ is vanishing, and the perturbation equation
reduces to
\beq
\ddot \psi_k +\left(k^2 +\a^2\eta^2\right)\psi_k=0,
\label{c6}
\eeq
where we have put
\beq
m^2a^2 =\a^2\eta^2, ~~~~~~~~~~~~~~~~
\a= mH_1a_1^2,
\eeq
using the time behaviour of the scale factor, $a \sim \eta$.

Assuming that the mass term is negligible at the transition scale, $
m \ll k/a$, we can match the solution (\ref{c5}) to the plane-wave
solution
\beq
\psi_k= {1\over \sqrt k}\left[c_+(k) e^{-ik\eta}+
c_-(k) e^{ik\eta}\right] ,
\label{c8}
\eeq
and obtain:
\beq
c_\pm=\pm c(k) e^{\pm ik\eta}, ~~~~~~~~~
|c(k)|\sim (k/k_1)^{-\mu-1/2} .
\label{c9}
\eeq
(We are neglecting, for simplicity, numerical factors of order 1,
which are not very significant in view of the many approximations
performed. Their contribution will be included into an overall
numericalcoefficient in front of the final spectrum.) In the
relativistic regime, the amplified axion perturbation then takes the
form:
\beq
\sg({\bf k}, \eta) = {c({\bf k})\over a \sqrt k}\sin (k\eta),
\label{c10}
\eeq
used in Section \ref{III3} for the massless-axion case.

In the radiation era the proper momentum is red-shifted
with respect to the rest mass, and all axion modes tend to become
non-relativistic. When the mass term is no longer negligible, the
general solution of Eq. (\ref{c6}) can be written in terms of 
parabolic cylinder  functions \cite{11}. For an approximate estimate
of the axion field in the non-relativistic regime, however, it is
convenient to distinguish two cases, depending on whether a mode
$k$ becomes non-relativistic inside or outside the horizon.  Defining
as $k_m$ the limiting comoving frequency of a mode that becomes
non-relativistic ($k_m=ma_m$) at the time it re-enters the horizon
($k_m=H_ma_m$), we find, in the radiation era,
\beq
k_m= k_1 \left(m\over H_1\right)^{1/2}.
\label{c11}
\eeq
We will thus consider the two cases $k \gg k_m$ and $k\ll k_m$.

In the first case, we rewrite the perturbation equation (\ref{c6}) as\beq
{d^2\psi_k\over dx^2}+\left({x^2\over 4} -b\right)\psi_k=0,
~~~~~ x=\eta (2 \a)^{1/2}, ~~~~ -b= k^2/2\a , \label{c12}
\eeq
and we give the general solution in the form
\beq
	\psi=A W(b,x)+B W(b,-x)~,\label{c13}
\eeq
where $W(b,x)$ are the Weber parabolic cylinder functions
(see \cite{11}, chap.~19). In order to fix the integration constants
$A$ and $B$ we shall match the solutions (\ref{c13}) and (\ref{c10})
in the relativistic limit
\beq
{k^2 \over m^2 a^2}=
{k^2 \over \a^2 \eta^2}= {-4b\over x^2} \gg1.
\label{c14}
\eeq
In this limit, as we are considering modes that become
non-relativistic when they are already inside the horizon,
\beq
\left(k\over k_m\right)^2 \sim {k^2\over \a} \sim (-b) \gg1,
\label{c15}
\eeq
we can expand the $W$ functions for $b$ large with $x$ moderate
\cite{11}. Matching to the plane-wave solution (\ref{c10}), we obtain
$A=0$, and
\beq
\psi_k \simeq {c({\bf k})\over \a^{1/4}} W(b,-x) .
\label{c16}
\eeq
In the opposite, non-relativistic limit $x^2 \gg |4b|$, the expansion
of the Weber functions gives \cite{11}
\beq
\psi_k \simeq {c({\bf k})\over (\a \eta)^{1/2}} \sin\left(m\over
H\right)
\label{c17}
\eeq
(we have used $x^2/4=ma\eta/2\sim m/H$). The corresponding 
axion field is (inside the horizon)
\beq
\sg({\bf k}, \eta) = {c({\bf k})\over a \sqrt{ma}}\sin \left(m\over
H\right), ~~~~~~~~~~~~~k>k_m .
\label{c18}
\eeq

Consider now the case of a mode that becomes non-relativistic when it  
is still
outside the horizon,  $k\ll k_m$. In this case, we cannot use the
large $|b|$ expansion as $|b| <1$, and it is convenient to express the
general solution of Eq. (\ref{c12}) as
\beq
	\psi=A y_1(b,x)+B y_2(b,x)~,\label{c19}
\eeq
where $y_1$ and $y_2$ are the even and odd parts of the parabolic
cylinder functions \cite{11}. Matching to (\ref{c10}), in the
relativistic limit $x \ra 0$, gives $A=0$ and
\beq
\psi_k \simeq c({\bf k})\left(k\over 2\a \right)^{1/2} y_2(b,x) .
\label{c20}
\eeq
In the non-relativistic limit $x^2 \gg |b|$ we use the relation
\cite{11}
\beq
y_2 \sim \left[W(b,x)-W(b,-x)\right] \sim {1\over \sqrt x} \sin
{x^2\over 4} ,
\label{c21}
\eeq
which leads to
\beq
\psi_k \simeq {c({\bf k})\over (\a \eta)^{1/2}} \left(k^2\over \a
\right)^{1/4} \sin\left(m\over H\right) .
\label{c22}
\eeq
Using Eqs.~(\ref{c15}) and (\ref{c11}) for $k^2/\a$, we finally arrive
at the non-relativistic axion field presented in Eq. (\ref{414}):
\beq
\sg({\bf k}, \eta) = {c({\bf k})\over a \sqrt{ma}}
\left(k\over k_1\right)^{1/2}\left(H_1\over m \right)^{1/4}
\sin \left(m\over H\right), ~~~~~~~~~~~~~k<k_m .
\label{c23}
\eeq

For later use, it is also convenient to define the spectral energy
density in critical units, $\Om_\sg (\om) = d(\r/\r_c)/d\ln \om$,
associated with the stochastic axion background in the three different
regimes defined before.

For relativistic modes we find, from Eq.
(\ref{c10}),
\beq
\Om_\sg (\om) \sim g_1^2 \left(\om\over \om_1\right)^{3-2\mu}
\left(H_1\over H\right)^2\left(a_1\over a\right)^4 ,
~~~~~~~~~ m<\om<\om_1.
\label{c24}
\eeq
For modes that becomes non-relativistic after re-entry we find, from Eq.
(\ref{c18}),
\beq
\Om_\sg (\om) \sim g_1^2 {m\over H_1}
\left(\om\over \om_1\right)^{2-2\mu}
\left(H_1\over H\right)^2\left(a_1\over a\right)^3 ,
~~~~~~~~~ \om_m<\om<m.
\label{c25}
\eeq
For modes that becomes non-relativistic before re-entry we find, from Eq.
(\ref{c23}),
\beq
\Om_\sg (\om) \sim g_1^2 \left(m\over H_1\right)^{1/2}
\left(\om\over \om_1\right)^{3-2\mu}
\left(H_1\over H\right)^2\left(a_1\over a\right)^3 ,
~~~~~~~~~ \om<\om_m.
\label{c26}
\eeq
The last two spectral distributions are constant during the
matter-dominated era, and the last one corresponds to the
spectrum of the Bardeen potentials, as given in Eq. (\ref{420}).

\end{document}